\documentclass[sigconf]{acmart}

\usepackage{type1cm}     
\usepackage{graphicx}     
\usepackage{xspace}     
\usepackage{balance}     
\usepackage{booktabs}     
\usepackage{multirow}      
\usepackage[tableposition=top]{caption}     
\usepackage[vlined,linesnumbered,ruled,noend]{algorithm2e}     
\usepackage{natbib}     
\usepackage{microtype}    
\usepackage{units}     
\usepackage{mathtools}     
\usepackage{amssymb}     
\usepackage{xfrac}    
\usepackage{float}
\usepackage[hide]{notes}
\usepackage{enumitem}
\usepackage{makecell}
\usepackage{subcaption}
\usepackage{flushend}
\usepackage{mathtools,xparse}
\usepackage{hyperref}

\graphicspath{{img/}}

\newcommand{\usergraph}{\ensuremath{G}\xspace}
\newcommand{\users}{\ensuremath{U}\xspace}
\newcommand{\user}{\ensuremath{u}\xspace}
\newcommand{\edges}{\ensuremath{E}\xspace}
\newcommand{\edge}{\ensuremath{e}\xspace}
\newcommand{\edgeweight}{\ensuremath{w}\xspace}
\newcommand{\nousers}{\ensuremath{n}\xspace}
\newcommand{\sources}{\ensuremath{S}\xspace}
\newcommand{\source}{\ensuremath{s}\xspace}
\newcommand{\nosources}{\ensuremath{m}\xspace}
\newcommand{\noclusters}{\ensuremath{k}\xspace}
\newcommand{\ideology}{\ensuremath{i}\xspace}
\newcommand{\UUmat}{\ensuremath{\mathbf{A}}\xspace}
\newcommand{\USmat}{\ensuremath{\mathbf{C}}\xspace}
\newcommand{\UCmat}{\ensuremath{\mathbf{U}}\xspace}
\newcommand{\SCmat}{\ensuremath{\mathbf{V}}\xspace}
\newcommand{\CCUmat}{\ensuremath{\mathbf{H}_u}\xspace}
\newcommand{\CCSmat}{\ensuremath{\mathbf{H}_s}\xspace}
\newcommand{\ULmat}{\ensuremath{\mathbf{L}_u}\xspace}

\newcommand{\SLmat}{\ensuremath{\mathbf{L}_s}\xspace}
\newcommand{\WLmat}{\ensuremath{\mathbf{L}_w}\xspace}
\newcommand{\ZLmat}{\ensuremath{\mathbf{L}_z}\xspace}
\newcommand{\regparamone}{\ensuremath{\alpha}\xspace}
\newcommand{\regparamtwo}{\ensuremath{\beta}\xspace}
\newcommand{\trace}[1]{\ensuremath{\mathit{tr}({#1})}\xspace}
\newcommand{\Dmat}{\ensuremath{\mathbf{D}}\xspace}
\newcommand{\Xmat}{\ensuremath{\mathbf{X}}\xspace}
\newcommand{\Wmat}{\ensuremath{\mathbf{W}}\xspace}
\newcommand{\Hmat}{\ensuremath{\mathbf{H}}\xspace}
\newcommand{\Zmat}{\ensuremath{\mathbf{Z}}\xspace}
\newcommand{\Umat}{\ensuremath{\mathbf{U}}\xspace}
\newcommand{\Lmat}{\ensuremath{\mathbf{L}}\xspace}
\newcommand{\eye}{\ensuremath{\mathbf{I}}\xspace}
\newcommand{\xrow}{\ensuremath{\mathbf{x}}\xspace}
\newcommand{\urow}{\ensuremath{\mathbf{u}}\xspace}
\newcommand{\usera}{\ensuremath{u}\xspace}
\newcommand{\userb}{\ensuremath{v}\xspace}

\newcommand{\useri}{\ensuremath{u_i}\xspace}
\newcommand{\userj}{\ensuremath{u_j}\xspace}
\newcommand{\userk}{\ensuremath{u_k}\xspace}
\newcommand{\obji}{\ensuremath{o_i}\xspace}
\newcommand{\objj}{\ensuremath{o_j}\xspace}
\newcommand{\objk}{\ensuremath{o_k}\xspace}
\newcommand{\objects}{\ensuremath{O}\xspace}
\newcommand{\rown}[1]{\ensuremath{{#1}_{\mathit{rn}}}\xspace}
\newcommand{\coln}[1]{\ensuremath{{#1}_{\mathit{cn}}}\xspace}
\newcommand{\xcoord}{\ensuremath{x}\xspace}
\newcommand{\ycoord}{\ensuremath{y}\xspace}
\newcommand{\iscore}[2]{\ensuremath{i({#1},{#2})}\xspace}
\newcommand{\popscore}[2]{\ensuremath{\rho({#1},{#2})}\xspace}

\DeclarePairedDelimiter{\abs}{\lvert}{\rvert}
\DeclarePairedDelimiter{\norm}{\lVert}{\rVert}

\DeclarePairedDelimiter{\vectorBrackets}{\langle}{\rangle}

\NewDocumentCommand{\normL}{ s O{} m }{%
    \IfBooleanTF{#1}{\norm*{#3}}{\norm[#2]{#3}}_{F}^2%
}
\NewDocumentCommand{\normF}{ s O{} m }{%
     \IfBooleanTF{#1}{\norm*{#3}}{\norm[#2]{#3}}_{F}%
}
\NewDocumentCommand{\normsquare}{ s O{} m }{%
    \IfBooleanTF{#1}{\norm*{#3}}{\norm[#2]{#3}}^2%
}

\newcommand{\reals}{\ensuremath{\mathbb{R}}\xspace}
\newcommand{\nmf}{{\sc\Large nmf}\xspace}
\newcommand{\NMFSymm}{{\sc\Large nmf-symm}\xspace}
\newcommand{\onmft}{{\sc\Large onmtf}\xspace}
\newcommand{\dmcc}{{\sc\Large dmcc}\xspace}
\newcommand{\prob}{{\sc\Large ifd}\xspace}
\newcommand{\probngr}{{\sc\Large ifd-ngr}\xspace}
\newcommand{\purity}{{\sc\Large purity}\xspace}
\newcommand{\AMI}{{\sc\Large ami}\xspace}
\newcommand{\NMI}{{\sc\Large nmi}\xspace}
\newcommand{\ARI}{{\sc\Large ari}\xspace}
\newcommand{\PMCC}{{\sc\Large corr}\xspace}
\newcommand{\PMCCi}{{\sc\Large corr$_i$}\xspace}
\newcommand{\PMCCp}{{\sc\Large corr$_\rho$}\xspace}
\newcommand{\kulshrestha}{{\sc\Large kulshrestha}\xspace}
\newcommand{\barbera}{{\sc\Large barbera}\xspace}
\newcommand{\retweet}{{\sc\Large retweet}\xspace}
\newcommand{\follow}{{\sc\Large follow}\xspace}
\newcommand{\avgcontent}{{\sc\Large avg\_content}\xspace}
\newcommand{\content}{{\sc\Large content}\xspace}
\newcommand{\biaswatch}{{\sc\Large biaswatch}\xspace}
\newcommand{\kde}{{\sc\Large kde}\xspace}

\newcommand{\fullprob}{{ideology factor decomposition}\xspace}

\newcommand{\spara}[1]{\smallskip\noindent\textbf{#1}}

\newcommand{\para}[1]{\noindent\textbf{#1}}

\newtheorem{problem}{Problem}

\newenvironment{squishlist}
{\begin{list}{$\bullet$}
  { \setlength{\itemsep}{0pt}
     \setlength{\parsep}{1pt}
     \setlength{\topsep}{1pt}
     \setlength{\partopsep}{0pt}
     \setlength{\leftmargin}{1.5em}
     \setlength{\labelwidth}{1em}
     \setlength{\labelsep}{0.5em} } }
{\end{list}}

\setcopyright{acmlicensed}
\copyrightyear{2018}
\acmYear{2018}
\acmConference[WSDM 2018]{WSDM 2018: The Eleventh ACM International Conference on Web Search and Data Mining }{February 5--9, 2018}{Marina Del Rey, CA, USA}
\acmPrice{15.00}
\acmDOI{10.1145/3159652.3159669}
\acmISBN{978-1-4503-5581-0/18/02}
\begin{document}

\title{Joint Non-negative Matrix Factorization for\\Learning Ideological Leaning on Twitter}


\author{Preethi Lahoti}
\authornote{Author was at Aalto University at the time of this work.}
\affiliation{
	\institution{Max Planck Institute for Informatics}
	\city{Saarbruecken}
	\country{Germany}
}
\email{plahoti@mpi-inf.mpg.de}

\author{Kiran Garimella}
\affiliation{
\institution{Aalto University}
\city{Helsinki}
\country{Finland}
}
\email{kiran.garimella@aalto.fi}

\author{Aristides Gionis}
\affiliation{
\institution{Aalto University}
\city{Helsinki}
\country{Finland}
}
\email{aristides.gionis@aalto.fi}


\begin{abstract}
People are shifting from traditional news sources to online news at an incredibly fast rate.
However, the technology behind online news consumption promotes content that confirms
the users' existing point of view.
This phenomenon has led to polarization of opinions and intolerance towards opposing views.
Thus, a key problem is to model information filter bubbles on social media
and design methods to eliminate them.  
In this paper, we use a machine-learning approach to learn a 
\emph{liberal-conservative} ideology space on Twitter, and 
show how we can use the learned latent space to tackle the filter bubble problem.

We model the problem of learning the \emph{liberal-conservative} ideology space 
of social media users and media sources
as a constrained non-negative matrix-factorization problem.
Our model incorporates the social-network structure and content-consumption information 
in a joint factorization problem with shared latent factors. 
We validate our model and solution on a real-world Twitter dataset consisting of 
controversial topics, 
and show that we are able to separate users by ideology with over $90$\,\% \emph{purity}. 
When applied to media sources, 
our approach estimates ideology scores that are highly correlated (Pearson correlation $0.9$) 
with ground-truth ideology scores. 
Finally, we demonstrate the utility of our model in real-world scenarios, 
by illustrating how the learned ideology latent space 
can be used to develop exploratory and interactive interfaces 
that can help users in diffusing their information filter bubble. 
\end{abstract}

\settopmatter{printacmref=false, printfolios=false}

\maketitle

\section{Introduction}
\label{section:introduction}


Social media and the web have provided a foundation where users 
can easily access diverse information from around the world. 
However, over the years, various factors, such as user homophily 
(social network structure), and algorithmic filtering 
(e.g., news feeds and recommendations) have narrowed the content that a user consumes. 
%
%
As an example, imagine two users of opposite ideological stances (liberal and conservative).
Though the two users may be looking at the same topic (e.g., a presidential debate), 
they might be seeing completely different viewpoints due to the diverse network surrounding 
and different content sources they get their information from.
Consequently, users on different ends of the ideological spectrum 
live in their own information bubbles~\cite{pariser2011filter}, 
oblivious to the views on the other side and creating their own world-view of truth.
This phenomenon has led to the polarization of viewpoints, 
intolerance towards opposing views, 
and ideological segregation~\cite{sunstein2009republic}. 
Many studies suggest that increasingly users live in their echo chambers~\cite{prior2007post} 
and polarization of the public has intensified~\cite{dimock2014political}.
%


In this paper, we propose a principled approach 
to infer the {\em ideological stances} 
(also known as {\em ideology} or {\em polarity} or {\em leaning}) 
of both the {\em users} in a social network, 
and the {\em media sources} that provide news content in the network.
Our approach is based on a 
non-negative matrix-factorization model, 
which jointly decomposes the social network of users and the content they consume
in a {\em shared latent space}. 
%
Learning the ideological stances of social media users and content sources 
is an important step in building useful tools that can 
make users aware of
their informational bias and consequently, help in reducing those biases, thus 
mitigating
the increasing polarization in the society.

Existing approaches to identifying the ideological leaning of users either\,  
(i) require large amounts of manually annotated data~\cite{conover2011predicting,pennacchiotti2011democrats}; or 
(ii) consider only the structure of social ties~\cite{barbera2015birds} 
or user interactions~\cite{wong2016quantifying,garimella2016quantifying}; or 
(iii) analyze only the content shared by users~\cite{pla2014political}. 
Each of these families of approaches has its own limitations. 
First, obtaining manual annotations is both expensive and time consuming.
Second, when using only the structure of social ties or interactions,  
it is assumed that users with the same ideology 
are more likely to interact with or follow each other. 
However, social networks are extremely sparse and noisy, 
and users with the same ideological leaning may never interact with or follow each other, 
while users with different ideological leaning may still interact due to different reasons.
Last, inferring ideological leaning by using only content is a challenging task
and prone to errors due to the inherent complexity of natural-language understanding.

To overcome these problems,
we propose an unsupervised approach that uses simultaneously the network structure and 
the information about content shared by users. 
We are motivated by the observation that a user's ideological stance on a topic 
depends on both the surrounding network structure as well as 
the content sources that users consume their information from.
Thus, we exploit the inherent connection between the two data types: 
Users are not only more likely to interact with or follow like-minded users, 
but also to share content of aligned ideological leaning. 
Taking into account the full available information 
results in more accurate estimation of ideological leaning 
both for users and content sources. 
%

In addition, we formulate the task of learning ideological leaning
as a joint matrix factorization problem
in which users and sources are represented in a shared latent space.
This formulation allows us to identify the relationship between data points of the two types, 
which is particularly useful for visualizing social-media users and sources in a common space, 
and building applications for exploring the ideological landscape
in one's media neighborhood, 
or making recommendations to escape the filter bubble.
In fact, we demonstrate concretely
how to use the learned ideology latent space 
to develop exploration and recommendation tools (Section~\ref{section:case-study}).

Experiments comparing our approach to the state of the art show the benefit of such a principled joint approach.
Our method is able to separate users into ideological clusters with over 90\% purity. 
When applied to media sources, our approach estimates ideology scores that are highly correlated with ground-truth ideology scores. 


All our analysis is done on Twitter, 
though the methods generalize to any other social network. 
In the rest of the paper, for the sake of clarity we use Twitter-specific nomenclature 
(e.g., retweets, follow, etc.).



In summary, our contributions are as follows:

\begin{squishlist}
\item  
We present a principled approach to jointly compute the ideology scores of both users and 
content sources by formulating the problem as a joint constrained matrix-factorization task. 
This formulation allows us to jointly cluster the two data types, ``user'' and ``source''.
We apply our learned ideological latent space
to the problems of ``who'' and ``what'' to recommend to users
so as to reduce polarization and diffuse their information filter bubble.
\item 
To the best of our knowledge, this is the first work to compute ideological stances 
for \emph{both} users and content sources, in a common latent space, 
based on both the network structure and the users' interaction with sources. 
\item 
We provide an extensive experimental evaluation, 
presenting both qualitative and quantitative results on real-world Twitter data.
Our method shows promise when compared to existing state of the art approaches.


\end{squishlist}










\section{Related work}\label{section:related-work}

In this paper, we propose an approach that can identify the ideological leaning of users on Twitter. We define ideology based on the policy dimension that articulates a user's political preference. 
Our definition is inspired by work in political-science literature, such as work by Bafumi et al.~\cite{bafumi2005practical}, where ideology is defined as ``a line whose left end is understood to reflect an extremely liberal position and whose right end corresponds to extreme conservatism.''

\para{Estimation of user ideology.}
Traditionally, the most common sources for estimating ideology comprised of behavioral data generated from 
roll call votes~\cite{poole2011ideology}, co-sponsorship records~\cite{aleman2009comparing}, or political contributions~\cite{bonica2013ideology}.
These datasets were often only available for the political elite, like members of congress, and hence getting such estimates for a large population of ordinary citizens was difficult, if not impossible.

With the proliferation of social media platforms, behavioral data started being available at an individual level and researchers have tried to use such data for identifying political ideology for social media users at scale.
Initial work started with supervised methods~\cite{pennacchiotti2011democrats,conover2011predicting} for predicting a (binary) political alignment of users on Twitter.
Though these works report accuracies over 90\%, Cohen at al.~\cite{cohen2013classifying} warns about the limitations of such approaches and their dependence on politically active users.

Unsupervised approaches have also been proposed, mainly based on the structure of user interests~\cite{kulshrestha2017quantifying}, social connections~\cite{barbera2015birds}, and interactions~\cite{bond2015quantifying,wong2016quantifying,garimella2016quantifying}.
The main idea behind these methods is that users typically either surround themselves (follow/friend) with other users who are similar in their ideology (homophily), or interact with others (retweet/like) similar to them.

Perhaps the closest approach to this paper is the work by Lu et al.~\cite{lu2015biaswatch} (\biaswatch), who seek to identify the {\em bias} of a user on a topic by combining their retweet and content networks, where a content network is obtained based on the similarity of users tweets. 
\biaswatch, however, assumes the presence of a set of labeled {\em bias anchors} (seed hashtags), 
making it  not completely unsupervised. 
Second, fusing the content and retweet networks is somewhat arbitrary, since there is no common underlying principle that holds the two networks together and hence a graph that results from such a merger contains different types of edges (multigraph) simply merged together.
We compared our approach with \biaswatch in Section~\ref{section:experiments} and show that our method outperforms their bias scores.


\para{Estimation of source ideology.}
Polarization and bias of media outlets has existed long before the time of internet, 
however, data availability makes it easier to study and quantify nowadays. 
Mitchell et al.~\cite{mitchell2014political} study the media habits of American public using a large scale survey and show how fractured the media production and consumption have become in the recent years.
Groseclosea et al.~\cite{groseclose2005measure} propose a method to estimate the ideology of various media sources 
by comparing the number of citations to think tanks and policy groups to those of Congress members.
For a complete survey of methods on measuring ideology of media and media bias, please refer to
Groeling et al.~\cite{groeling2013media}.


\para{Reducing polarization.}
The problem of reducing polarization and diffusing information bubbles on social media has been tackled before.
Most papers approach the problem by recommending content outside a user's filter bubble.
Studies have looked at what to recommend~\cite{graells2014people}, how to recommend~\cite{liao2014can,liao2014expert} and to whom to recommend such content~\cite{garimella2017reducing}.
%

Based on the papers discussed above, we make the following observations:

a. Most existing approaches only consider one dimension (content or network) for computing user ideology. We propose a principled approach that can compute a user's ideology by taking into account \emph{both} content and network. 

b. We compute the ideology of users and sources simultaneously. 
No existing approach, to the best of our knowledge, proposes such a method to do it simultaneously.

c. Most approaches for reducing polarization do not take a user's \emph{choice} into account. 
Our methods (in Section~\ref{section:case-study}) can help users diffuse their information bubbles based on their own choices.

\section{Problem formulation}
\label{section:problem}

In this section we discuss the problem setting, 
introduce the notation,
and formally define the problem we consider.

\subsection{Data model}

\spara{The social graph.}
We consider a social graph of Twitter users
$\usergraph=(\users,\edges,\edgeweight)$, 
where \users represents the set of users, 
\edges is the set of edges representing social interactions between the users,
and
$\edgeweight : \edges \rightarrow \reals$ is a weighting function
that assigns real-valued weights to each edge in \usergraph.

We also represent the graph \usergraph by its adjacency matrix 
$\UUmat\in\reals^{\nousers \times \nousers}$.
In particular, it is $\UUmat(\usera,\userb)=\edgeweight_i$
for each edge $\edge_i = (\usera,\userb,\edgeweight_i)\in \edges$, 
and $\UUmat(\usera,\userb)=0$ if there is no edge between \usera and \userb.
Since the social graph \usergraph may not represent a symmetric relation, 
the matrix \UUmat is not necessarily symmetric. 

To define the edges of the graph \usergraph
one can consider social links between users 
based on their retweet and follow networks. 
Two commonly-used options are to consider 
an edge $\edge_i = (\usera,\userb,\edgeweight_i)$ if
(i) users \usera and \userb \textit{follow} a common set of users, and 
$\edgeweight_i$ is the \textit{number of common users}; or if 
(ii) user \usera \textit{retweets} user \userb, and 
$\edgeweight_i$ is the \textit{number of retweets}. 
In the first case the social graph is symmetric, 
while in the second case it is not. 


\spara{Content sources.}
We model the presence of content in the network
by considering a set of content sources \sources.
The set \sources represents a set of items that are shared in the network. 
The number of sources is denoted by $\abs{\sources} =\nosources$.
We associate with each user \user the subset of sources in \sources 
that the user shares (i.e., posts) in the network. 
Thus the overall user activity in the network is denoted
by a matrix $\USmat\in\reals^{\nousers\times\nosources}$, 
where $\USmat(\user,\source)$ denotes the number of times that user $\user\in\users$ 
shares source $\source\in\sources$. 

We derive the content features (i.e., items of the set \sources)
based on users' tweets. 
Specifically, we experiment with two variants of feature sets: 
(i) extracted urls from the tweets; and 
(ii) hostnames of urls extracted from the tweets. 
In other words, in the latter case, 
the content is aggregated by their source of authorship 
(i.e., various news media channels).


\smallskip
For the presentation of our method, 
we also need the following definitions.

\spara{Affinity matrix.}
\label{section:affinity}
Given a matrix \Xmat, 
we define the {\em affinity matrix of the rows of}\, \Xmat to be  
$\rown{\Xmat}\rown{\Xmat}^T$,
where  $\rown{\Xmat}$ is row normalized \Xmat. 
That is, the affinity matrix of rows of \Xmat is 
formed by the cosine similarities of all pairs of rows of \Xmat. 
Similarly, we define the {\em affinity matrix of the columns of}\, \Xmat 
to be $\coln{\Xmat}^T\coln{\Xmat}$, 
where $\coln{\Xmat}$ is column normalized~\Xmat. 

\spara{Laplacian matrix.}
\label{section:affinity}
Given a square and symmetric matrix \Xmat, 
we consider its Laplacian to be the matrix $\Lmat = \Dmat - \Xmat$,
where \Dmat is a diagonal matrix whose $(i,i)$ entry
is the $i$-th row-sum of~$\Xmat$.



\subsection{Motivation of the approach}



Our goal in this paper is to 
learn the ideological leaning of users 
and content sources 
on Twitter.
The underlying motivation 
is that learning the ideological leaning of Twitter users and sources
is an important step in building useful tools that can help users  
perceive their informational bias and consequently 
improve their news diet.

A simple approach to identify ideological leaning of users is 
to consider only the social graph
and apply one of the many community-detection algorithms. 
The intuition here is that users with the same ideological leaning
are more likely to interact with each other, 
and thus, to form graph communities.
The drawback of this simple approach is that 
community detection on real-world social graphs is an extremely challenging task, 
due to sparsity and overlapping communities. 
For instance, two users \usera and \userb
can have the same ideological leaning 
even though they do not have any social interaction, 
or they have different topical interests. 

Similarly, 
one simple approach to attempt identifying the ideological leaning of sources is 
to cluster them using deep-NLP or semantic-analysis techniques.
However, this approach is again prone to errors due to the inherent complexity
of the text-analysis task.
Furthermore, it ignores the rich user information 
about how content is shared in the social network.

In contrast to these simple techniques, 
which rely on one-sided clustering of either users or content, 
with no association between them, 
the proposed approach seeks to 
{\em jointly} learn the ideological leaning of users and content sources. 

Combining different data types in a unified learning framework has several advantages. 
First, we exploit the inherent connection between the two data types: 
users are not only more likely to interact with, or follow, like-minded users, 
but also to share content of aligned ideological leaning.
Consequently, taking into account the full information on social structure and content
will result in better clustering performance. 
Furthermore, considering both data types simultaneously
allows us to learn the ideological leanings of users and content sources
in a {\em shared latent space}. 
This means that not only do we separate users and sources into ideological clusters, 
but we also identify the {\em relationship} between the clusters of the two data types.
The applications we present in the Section~\ref{section:case-study}
rely heavily on the ability to represent users and content sources in a shared latent space.

\subsection{Problem formulation}
\label{section:problem-formulation} 

To learn the latent space of the input data (users and sources)
we use non-negative matrix factorization (\nmf) techniques.
In particular, we propose a joint matrix factorization formulation, 
which exploits the duality between user and source clustering. 

First, assume that the ideological leaning is represented by \noclusters factors (dimensions). 
Our model is described by two components. 
The first component, represented by a $\nousers\times\noclusters$ matrix \UCmat, 
captures user information. 
In particular, the entry $(\user,\ideology)$ of the matrix \UCmat 
represents the degree to which user \user aligns with 
ideology factor \ideology.

The second component, represented by a $\nosources\times\noclusters$ matrix \SCmat, 
captures source information: 
the entry $(\source,\ideology)$ of the matrix \SCmat 
represents the degree to which source item \source aligns with 
ideology factor \ideology.

To determine the user and source ideology clusters, 
we decompose the two input matrices  \UUmat and \USmat, 
using the components \UCmat and \SCmat as latent factors. 
For decomposing \UUmat and \USmat so as to capture ideology clusters
we require the following constraints. 


\spara{Partitioning constraints:} 
\begin{squishlist}
\item[1.] Users in a user-cluster interact with each other more often than with users outside the cluster.
\item[2.] Users in the same user-cluster post content from the same content-cluster.
\item[3.] Content in the same content-cluster is posted by users in the same user-cluster.
\end{squishlist}
\para{Co-partitioning constraints:} 
\begin{squishlist}	
\item[4.] Users in a user-cluster share more articles from their corresponding content-cluster 
than from other content-clusters.
\item[5.] Content in a content-cluster is shared by more users from their corresponding user-cluster 
than from other user-clusters. 
\end{squishlist}


\spara{Non-negative matrix factorization for co-clustering:}
For a given input data matrix \Xmat, 
the bi-orthogonal non-negative 3-factor decomposition (\onmft)~\cite{ding2006orthogonal}, 
formulated as $\Xmat \approx \Wmat \Hmat \Zmat^T$,
provides a good framework for simultaneously clustering the rows and the columns of~\Xmat.
The left factor  \Wmat provides a clustering of the rows of \Xmat, 
while the right  matrix \Zmat provides a clustering of the columns of~\Xmat.
The middle factor \Hmat provides association between the clusters 
and additional degrees of freedom. 
\begin{problem}
\label{problem:onmft}
{\em (}\onmft~\cite{ding2006orthogonal}{\em )} 
Given an $n \times m$ matrix \Xmat, 
and integer $k$, with $k << n,m$,
find non-negative matrices \Wmat, \Hmat, and \Zmat, 
of respective dimensions
$n \times k$, 
$k \times k$, and
$k \times m$,
so as to 
\begin{align}
\label{equation:onmft}
\textrm{minimize }\; & \normF{\Xmat - \Wmat \Hmat \Zmat^T}^2, \nonumber \\
\textrm{subject to }\; & \Wmat\geq 0, \Hmat\geq 0, \Zmat\geq 0, \nonumber \\
\textrm{and }\; & \Wmat^T\Wmat = \Zmat^T\Zmat = \eye. \nonumber 
\end{align}
\end{problem}
As shown by Ding et al.~\cite{ding2006orthogonal}, 
the bi-orthogonality constraints 
provide an interpretation of the \onmft problem
as a simultaneous clustering of the rows and columns of the input data matrix.

In our approach we use \onmft to decompose the input matrices
(social-graph matrix \UUmat and user--source matrix \USmat)
using the latent factors \UCmat and \SCmat.

First, the user--source matrix is decomposed by
$\USmat \approx \UCmat \CCSmat \SCmat^T$, 
subject to orthogonality constraints for \UCmat and \SCmat.
%

For the social-graph matrix \UUmat, 
as both the rows and columns represent users,
we equate the left and right factors
and require  
$\UUmat \approx \UCmat \CCUmat \UCmat^T$, 
subject to orthogonality constraints for \UCmat.

Note that as \CCUmat is not necessarily symmetric, 
the decomposition $\UCmat \CCUmat \UCmat^T$
can produce a non-symmetric matrix.
Furthermore, the formulation can capture
link transitivity~\cite{zhu2007combining}:
consider a path
$\useri \rightarrow \userk \rightarrow \userj$. 
A non-symmetric factorization 
$\UUmat \approx \Xmat \Zmat^T$ represents the values of \UUmat 
as links from set of users to a set of objects,
say, $\objects = \{\obji\}$. 
Hence, it would split the path 
($\useri \rightarrow \userk \rightarrow u_j$) 
into two parts $\useri \rightarrow \objk$ and $\userk \rightarrow \objj$, 
which is a misinterpretation of the original path. 
Whereas, $\UUmat \approx \UCmat \CCUmat \UCmat^T$ 
considers the links to be amongst the same set of objects. 
Hence, the transitive link 
$\useri \rightarrow \userk \rightarrow u_j$ is correctly captured by 
the latent factors in \UCmat.


\spara{Combining link and content information}
is achieved by using a common latent factor \UCmat
and formulating a joint factorization problem 
asking to minimize 
\begin{equation}
\label{equation:link+content}
\normF{\UUmat - \UCmat \CCUmat \UCmat^T}^2 +
\normF{\USmat - \UCmat \CCSmat \SCmat^T}^2,
\end{equation}
subject to non-negativity of 
\UCmat, \CCUmat, \SCmat, \CCSmat, 
and orthogonality of 
\UCmat and \SCmat. 
This approach is inspired by the formulation of 
Zhu et al.~\cite{zhu2007combining} for classifying web-pages by exploiting both content and link information.

\spara{Graph regularization.}
From a geometric perspective, a dataset can be viewed as a set of data points on a continuous manifold. 
The task of clustering is to find these intrinsic manifolds in the data. 
However, the clustering formulation of \onmft 
fails to consider this geometric structure in the data.
To address this problem, Cai et al.~\cite{cai2008non} introduced a graph-regularized \nmf 
based on the \textit{manifold assumption} that if two data points 
$\xrow_i$, $\xrow_j$ are close in the input data matrix \Xmat, 
their projections
$\urow_i$ and $\urow_j$ in the new basis \Umat 
are also close. 
This is formulated by seeking to minimize
\[
\frac{1}{2}\sum\limits_{i,j}\normsquare{\urow_i - \urow_j}\Wmat_{ij} = 
\trace{\Umat^T \Lmat \Umat}, 
\]
where \Wmat is the affinity matrix of rows of \Xmat
(see Section~\ref{section:affinity}), 
and \Lmat is the laplacian of \Wmat.


Motivated by the duality between row and column manifolds, 
Gu et al.~\cite{gu2009co} proposed a dual-manifold regularized co-clustering 
method as a decomposition of an input matrix \Xmat, 
by asking to minimize
\[
\normF{\Xmat - \Wmat \Hmat \Zmat^T}^2
+ \alpha\cdot\trace{\Wmat^T \WLmat \Wmat} + 
\beta\cdot\trace{\Zmat^T \ZLmat \Zmat}, 
\]
subject to non-negativity constraints of the factor matrices. 
Here \WLmat and \ZLmat are the Laplacians on the affinity matrices 
of rows and columns of \Xmat, respectively. As before, the matrix \Hmat 
captures the association between row and column clusters.

We extend our joint factorization model (Equation~\ref{equation:link+content})
by including dual-graph regularization constraints on users and sources.
We refer to this problem by \prob, 
for {\em\fullprob}.
The formal problem definition is the following. 

\begin{problem}[\prob]
\label{problem:x}
Given a user--user social matrix \UUmat of dimension $\nousers \times \nousers$, 
a user--source matrix \USmat of dimension $\nousers \times \nosources$,
an integer \noclusters, with $\noclusters << \nousers,\nosources$, 
and regularization parameters \regparamone and \regparamtwo,
find factors 
\UCmat, \CCUmat, \SCmat, \CCSmat, 
of dimensions
$\nousers \times \noclusters$, 
$\noclusters \times \noclusters$, 
$\nosources \times \noclusters$, 
$\nousers \times \noclusters$, 
respectively, 
so as to 
\begin{align}
\label{equation:x}
\textrm{minimize }\; & 
	\normF{\UUmat - \UCmat \CCUmat \UCmat^T}^2 
	+ \normF{\USmat - \UCmat \CCSmat \SCmat^T}^2 \nonumber \\
 & 	+ \regparamone\cdot\trace{\UCmat^T \ULmat \UCmat} 
    + \regparamtwo\cdot\trace{\SCmat^T \SLmat \SCmat}, \\
\textrm{subject to }\; & 
\UCmat\geq 0, \CCUmat\geq 0, \SCmat\geq 0, \CCSmat \geq 0, \nonumber \\
\textrm{and }\; & \UCmat^T\UCmat = \SCmat^T\SCmat = \eye, \nonumber 
\end{align}
where \ULmat and \SLmat are the Laplacians 
of the affinity matrices of rows (users) and columns (sources)
of \USmat, respectively. 
\end{problem}

In the \prob problem we factorize \UUmat and \USmat jointly 
based on the dual manifold assumption i.e., 
both users and content share the same latent space and 
the cluster labels of users are smooth with respect to the content manifold, 
while the cluster labels of content are smooth with respect to the user manifold. 
To apply these manifold constraints, we consider affinity matrices for users and content sources.
While there are many ways to construct such an affinity matrices, 
in our experiments we consider the affinity matrices of 
the rows (for users) and columns (for sources) of the \USmat matrix.

Finally, it is instructive to review how  \prob
addresses the {\em partitioning constraints} 1--3 and 
{\em co-partitioning constraints} 4--5:
the bi-orthogonality constraints on the tri-factorization problem 
provide a clustering 
interpretation for the rows and columns of the input matrices, 
while the manifold constraints provide a geometric interpretation 
to the discovered latent space. 
The correspondance between the user clusters and source clusters
is achieved by using a shared latent factor.


\subsection{Solving the optimization problem} 
\label{section:solution} 

Following the standard theory of constrained optimization, 
we solve Problem~\ref{problem:x} 
by introducing Lagrangian multipliers $\Lambda$ 
(a symmetric matrix of size $\noclusters \times \noclusters$) 
and minimizing the Lagrangian function
\begin{eqnarray}
\label{equation:optimization}
 L & = & \normF{\UUmat - \UCmat \CCUmat \UCmat^T}^2 
	     + \normF{\USmat - \UCmat \CCSmat \SCmat^T}^2 \nonumber \\
   &   & + \regparamone\cdot\trace{\UCmat^T \ULmat \UCmat} 
         + \regparamtwo\cdot\trace{\SCmat^T \SLmat \SCmat} \nonumber \\
   &   & \trace{\Lambda(\UCmat^T\UCmat - \eye)} + \trace{\Lambda(\SCmat^T\SCmat - \eye)}.
\end{eqnarray}
 
We can compute the gradient of $L$ with respect to 
\UCmat, \SCmat, \CCUmat, and~\CCSmat. 
A locally-optimal solution for Problem~\ref{problem:x} 
can be found using an iterative-update algorithm, 
similar to the one proposed by Ding et al.~\cite{ding2006orthogonal}. The multiplicative update rules are as follows:

\begin{align}\label{eq-update-rules}
	& U \leftarrow U \sqrt{\frac{AUH_1^T + CVH_3^T+\alpha S_{u} U}{UH_1U^TUH_1^T + UH_3V^TVH_3^T + \alpha D_u U + U \lambda_u}}, \\
	& V \leftarrow V \sqrt{\frac{C^TUH_3 +\beta S_{d} V}{ \beta D_d V + VH_3U^TUH_3 + V \lambda_v}},\\
	& H_1 \leftarrow H_1 \sqrt{\frac{U^TAU}{U^TUH_1U^TU}},\\
	& H_3 \leftarrow H_3 \sqrt{\frac{U^TCV}{U^TUH_3V^TV}},
\end{align}
where
\begin{align*}
	& \lambda_u = U^TAUH_1^T + U^TCVH_3^T - \alpha U^T L_{u} U - H_1U^TUH_1^T - H_3V^TVH_3^T \\
	& \mbox{and } \lambda_v = V^TC^TUH_3 - \beta V^T L_{d} V - H_3U^TUH_3.
\end{align*}

\section{Estimating ideological leaning}
\label{section:methodology}

In this section we discuss how to use the latent factors
\UCmat and~\SCmat
computed using~\prob
in order to estimate ideological leaning scores for Twitter users
and media channels (sources).
Our approach utilizes the probabilistic model of \nmf factorizations,
discussed in Yoo et al.~\cite{yoo2010orthogonal}, 
to derive a probabilistic interpretation of latent factors. 
We also discuss how to derive a hard cross-ideological separation 
(hard clustering) from the latent factors. 

\begin{figure}[t]
\setlength\belowcaptionskip{-5pt}
	\begin{subfigure}{0.49\columnwidth}
		\centering
		\includegraphics[scale=0.36]{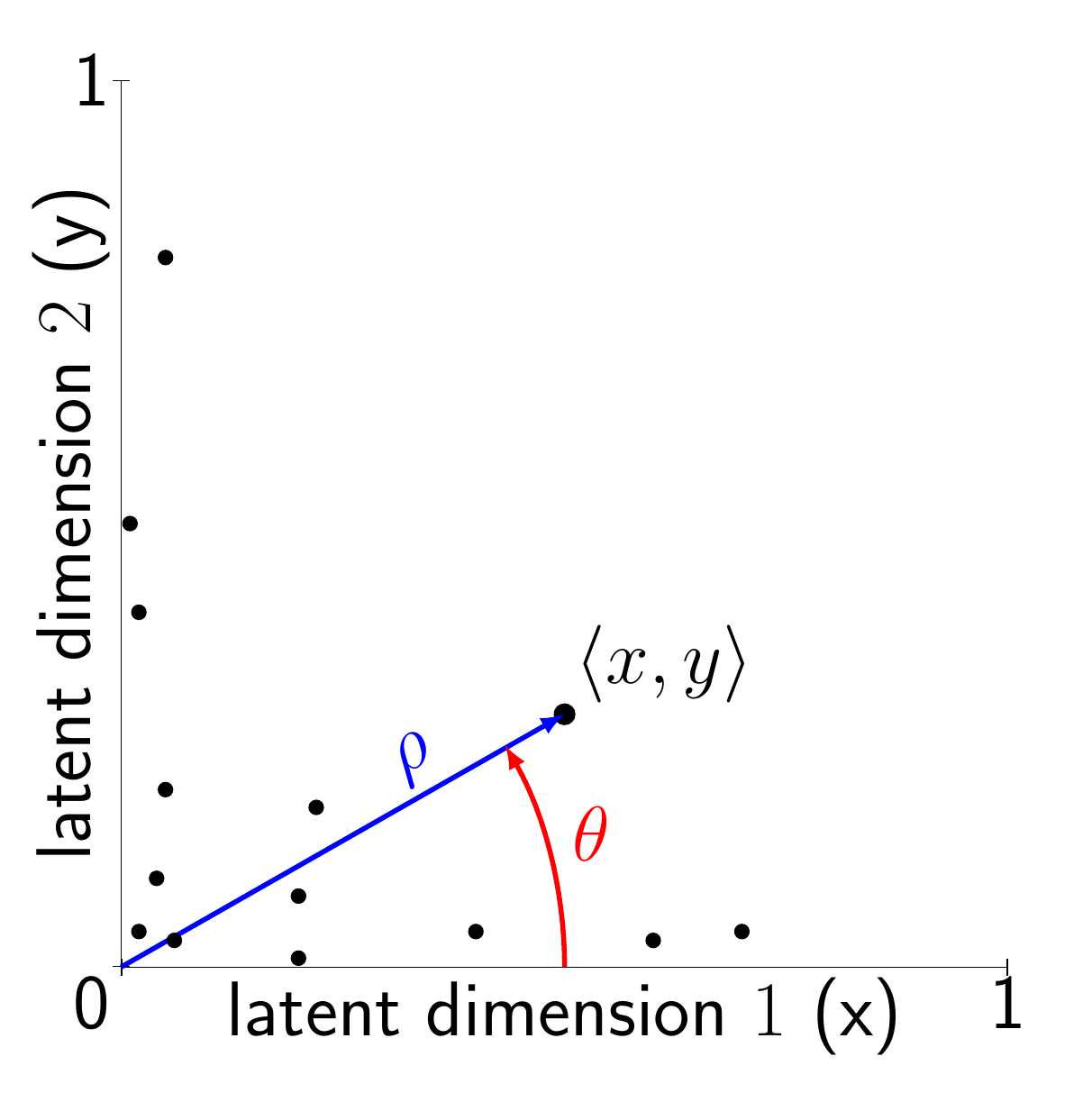}
		\caption{original latent space}
		\label{fig:latent-polar}
	\end{subfigure}
	\hfill
	\begin{subfigure}{0.5\columnwidth}
		\centering
		\includegraphics[scale=0.36]{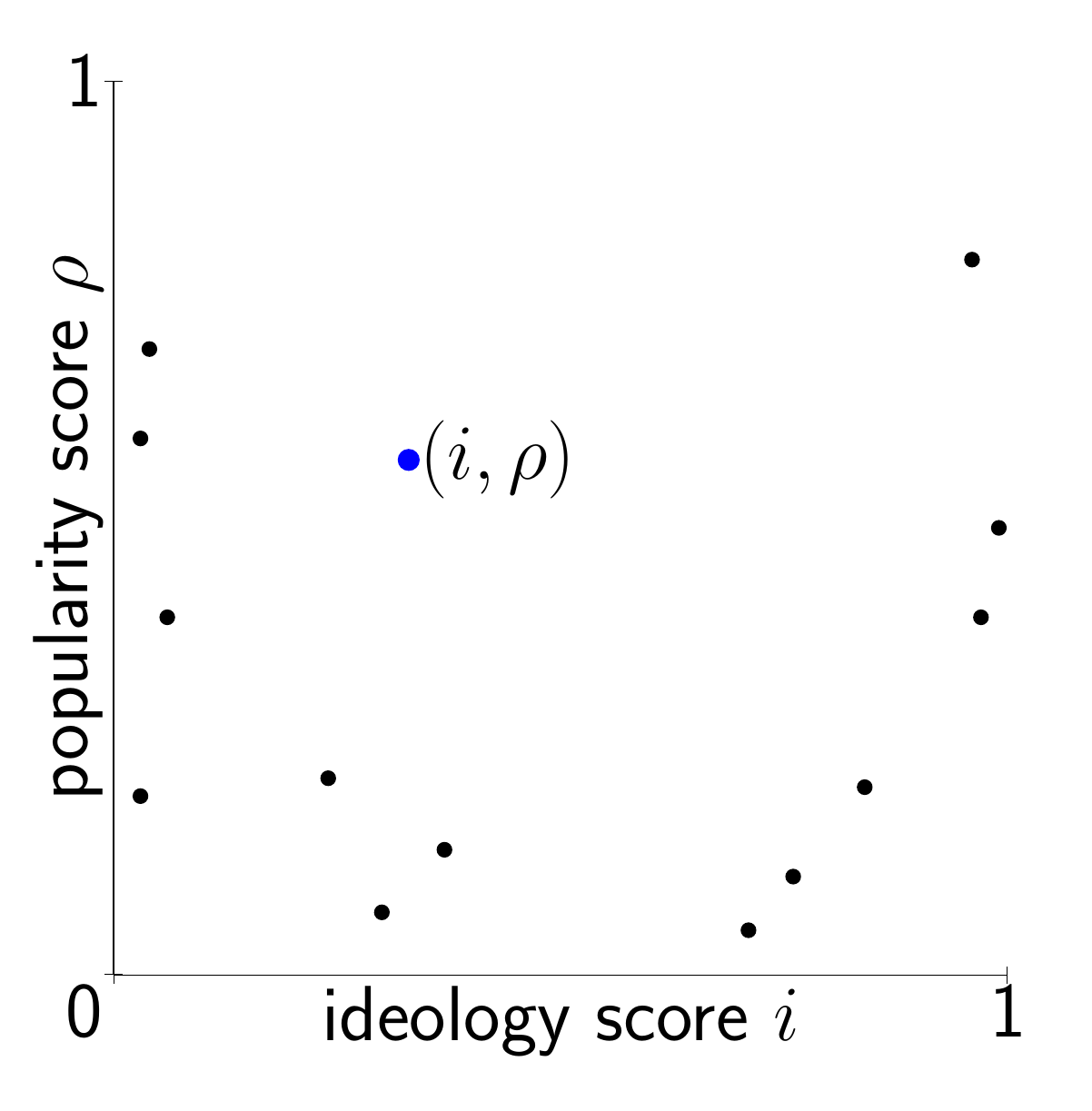}
		\caption{transformed coordinate space}
		\label{fig:latent-cartesian}
	\end{subfigure}
	\caption{\label{figure:latents}Projection of a subset data points in the learned ideology latent space 
	and the transformed ideology-popularity coordinate space.}
    \ctodo[Preethi]{Done. Review once}
    \ctodo[Kiran]{Can you also replace the $\theta$ with i(x,y)? and change the x and y axis labels to say 'latent dimension 1 (x)', and 'latent dimension 2 (y)'}
    \ctodo[Preethi]{i(x,y) is not equivalent to $\theta$. It is normalized $\theta$. (Equation 4.)}
\vspace{-\baselineskip}
\end{figure}

As proposed in the literature~\cite{li2006relationships,yoo2010orthogonal}, 
the latent factors \UCmat and \SCmat have the following probabilistic interpretation:
\begin{squishlist}
\item[--]
the entry $(i,\ell)$ of matrix \UCmat indicates the degree to which user $i$ 
belongs to the user-cluster $\ell$; and
\item[--]
the entry $(j,\ell)$ of matrix \SCmat indicates the degree to which media source $j$ 
belongs to the content-cluster $\ell$.
\end{squishlist}
In the context of our problem setting,
we are interested in identifying two main ideologies, 
\textit{liberal} and \textit{conservative}, 
and thus, we set the number of latent dimensions equal to $2$ ($\noclusters=2$). 
It follows that each user and each media source
are represented by a 2-dimensional vector $(\xcoord,\ycoord)$ in the latent space ---
a row of \UCmat for users, and a row of \SCmat for media sources. Thus, user and source ideology hard clusters are derived as $\arg\max U_{ij}$ and $\arg\max V_{ij}$, respectively.
To estimate a single score for users and media sources, we compute the angle of line defined by the center of origin
and the latent vector $(\xcoord,\ycoord)$, 
normalized to be between 0 and 1, i.e.,\footnote{Recall that 
$\arctan(0)=0$ and 
$\lim_{z\rightarrow\infty}\arctan(z)=\pi/2.$}
\begin{equation}
\label{equation:ideology-score}
\iscore{\xcoord}{\ycoord} = \frac{\theta}{\pi/2} = \frac{\arctan(\ycoord/\xcoord)}{\pi/2}.
\end{equation}
We also compute the magnitude of the latent vector $(\xcoord,\ycoord)$
\begin{equation}
\label{equation:popularity-score}
\popscore{\xcoord}{\ycoord} = \sqrt{{\xcoord}^2+{\ycoord}^2},
\end{equation}
which, as can be shown easily, 
represents the popularity of the corresponding user or source ---
in particular, it is correlated with the number of re-tweets and follows of a given user, 
and the number of tweets containing a given source. 
Figure \ref{figure:latents} visualizes a subset of real data points projected in the original latent space and their transformation to the corresponding ideology/popularity co-ordinate space according to the aforementioned computations.

In summary, given a user or media source 
represented by a latent vector $(\xcoord,\ycoord)$,
we can estimate a single 
ideological leaning score by Equation~(\ref{equation:ideology-score}), 
as well as its popularity by Equation~(\ref{equation:popularity-score}).
When using more than 2 dimensions for the latent ideology space ($\noclusters>2$)
it is not possible to estimate ideology scores with a single number, 
but we can still handle the user and source representations by 
standard vector operations.

\section{Experiments}\label{section:experiments}

\subsection{Dataset} 
Our dataset is collected using Twitter's streaming API from 
$2011$ to $2016$, 
by filtering for keywords related to three popular controversial topics: 
{\em gun control}, {\em abortion} and {\em obamacare}. 
We use the list of keywords proposed by Lu et al.~\cite{lu2015biaswatch} 
to filter the tweets related to these topics.
%
We only consider users who tweeted about all three topics at least once, 
obtaining a set of $\nousers=6\,391$ users, 
and collect all their tweets, which gave us 19 million tweets.

As discussed in Section~\ref{section:problem}, we consider two variants of matrix \UUmat 
(\emph{retweet} and \emph{follow}) and two variants of matrix \USmat (\emph{urls} and \emph{hostnames}). 
We observe that the \emph{follow} user-user matrix \UUmat 
along with \emph{hostname} user-content matrix \USmat 
gives significantly superior results compared to all other variants, 
thus, 
all subsequent results use that variant.
We omit results with the other variants due to lack of space.

\subsection{Ground truth} \label{sec-groundtruth}

\spara{Ideology scores for sources:} 
We collect ground truth for news-media channels from multiple studies in the literature: 
(i)  $500$ most shared news domains on Facebook~\cite{bakshy2015exposure}  
(ii) $100$ most visited domains in Bing toolbar~\cite{flaxman2016filter} and 
(iii) $27$ domains from an offline survey and webpage visit data~\cite{gentzkow2011ideological}. 
Each of these scores roughly measures the fraction of views/shares/clicks 
by a conservative user. 
We map all scores in the $[0,1]$ range, $1$ being conservative. 
For the domains listed in multiple lists we compute the ideology by averaging the scores. 
We also remove domains that are not necessarily news sources 
(e.g., {\em wikipedia.org}, {\em reddit.com}, etc.). 
In total, we collect $559$ news domains with ground-truth ideology scores.
We refer to this dataset as \content ground truth.
	
\spara{Ideology scores for Twitter users:} 
We use two different ground-truth scores for users: 
(i) \barbera: ideology score estimated by Barber{\'a} et al.~\cite{barbera2015tweeting}, 
which applies Bayesian ideal point estimate on nearly $12$ million Twitter users, and 
(ii) \avgcontent: average ground-truth ideology scores of the sources tweeted by the user.
	
	
\spara{Popularity scores for sources:} 
We use the aggregated number of tweets about each news media channel 
in the collected data set as a proxy for the popularity of the source. 

\spara{Popularity scores for users:} 
Since the collection of users is a random set of people on user, 
we do not have any ground truth for popularity of Twitter users.
\subsection{Baseline algorithms}

We compare our method with three types of methods for ideology detection: 
network-only, content-only, and a combination of network and content. 
%
%

\spara{Network-only:} We consider two types of network-only methods: 
(i) \nmf-based methods that can provide a continuous ideology score for a user between 0 and 1; and 
(ii) other methods that only produce binary labels for ideology (a user is either liberal or conservative). 
We use symmetric \nmf (\NMFSymm)~\cite{ding2006orthogonal}, 
a 3-factor \nmf shown to be equivalent to normalized-cut spectral clustering~\cite{ding2005equivalence}, 
\retweet a method based on partitioning the retweet graph~\cite{garimella2016quantifying} and \follow a graph partitioning approach on the follow network.  
In order to construct a source-source relationship matrix, we use $\USmat^T\USmat$. 
It is noteworthy that network-only methods perform only one-side clustering --- one data type at a time. 
Hence, we need to apply the methods separately for users and content sources. 
As such, network-only methods do not provide 
any information about the correspondence between the two clusterings. 
Further, \retweet and \follow return only binary labels, 
hence we do not use this baseline for comparing ideology scores.

\spara{Content-only:} 
We use orthogonal \nmf tri-factorization (\onmft), 
a co-clustering approach~\cite{ding2006orthogonal}, and 
dual manifold co-clustering (\dmcc)~\cite{gu2009co}. 
In these methods the bipartite content matrix \USmat is used to 
co-cluster the rows (users) and columns (sources) of the matrix simultaneously 
using bi-orthogonality and graph-regularization constraints.

\spara{Network and content:} 
We compute ideology scores of Twitter users 
estimated by kulshrestha et al.~\cite{kulshrestha2017quantifying} (\kulshrestha) and 
Lu et al.~\cite{lu2015biaswatch} (\biaswatch). 

\spara{Proposed methods:}
We use the proposed method \prob, 
and a variant of \prob without graph-regularization constraints (\probngr).
We initialized \UCmat and \SCmat randomly from a uniform distribution in [0,1] and
\CCUmat and \CCSmat as identity matrices of size \noclusters.
Parameters \regparamone and \regparamtwo are chosen using grid search.
Additional details on various approaches tried for parameter initialization, 
parameter tuning, and stability of the algorithms with respect to the parameters 
are omitted
due to lack of space, 
and will be provided in the full version of the paper.

\subsection{Experimental setup}

\spara{Evaluation measures.}
We perform two types of qualitative evaluation tasks: 
(i) quality of ideological cluster separation (into \emph{liberal} or \emph{conservative} clusters) and 
(ii) correlation between the computed ideology scores and ground-truth scores. 
In order to evaluate cluster separation, we measure \purity, 
adjusted Rand index (\ARI), 
adjusted mutual information (\AMI), and 
normalized mutual information (\NMI) 
between the clusters detected by the algorithm 
and the set of ground-truth communities derived by separating users at ideology score threshold at $0.5$.
In order to measure correlation between the computed ideology scores and the ground-truth scores,
we use Pearson mutual correlation coefficient (\PMCC).

\subsection{Results}

\begin{figure}[t]
	\centering	
	\includegraphics[scale=0.4]{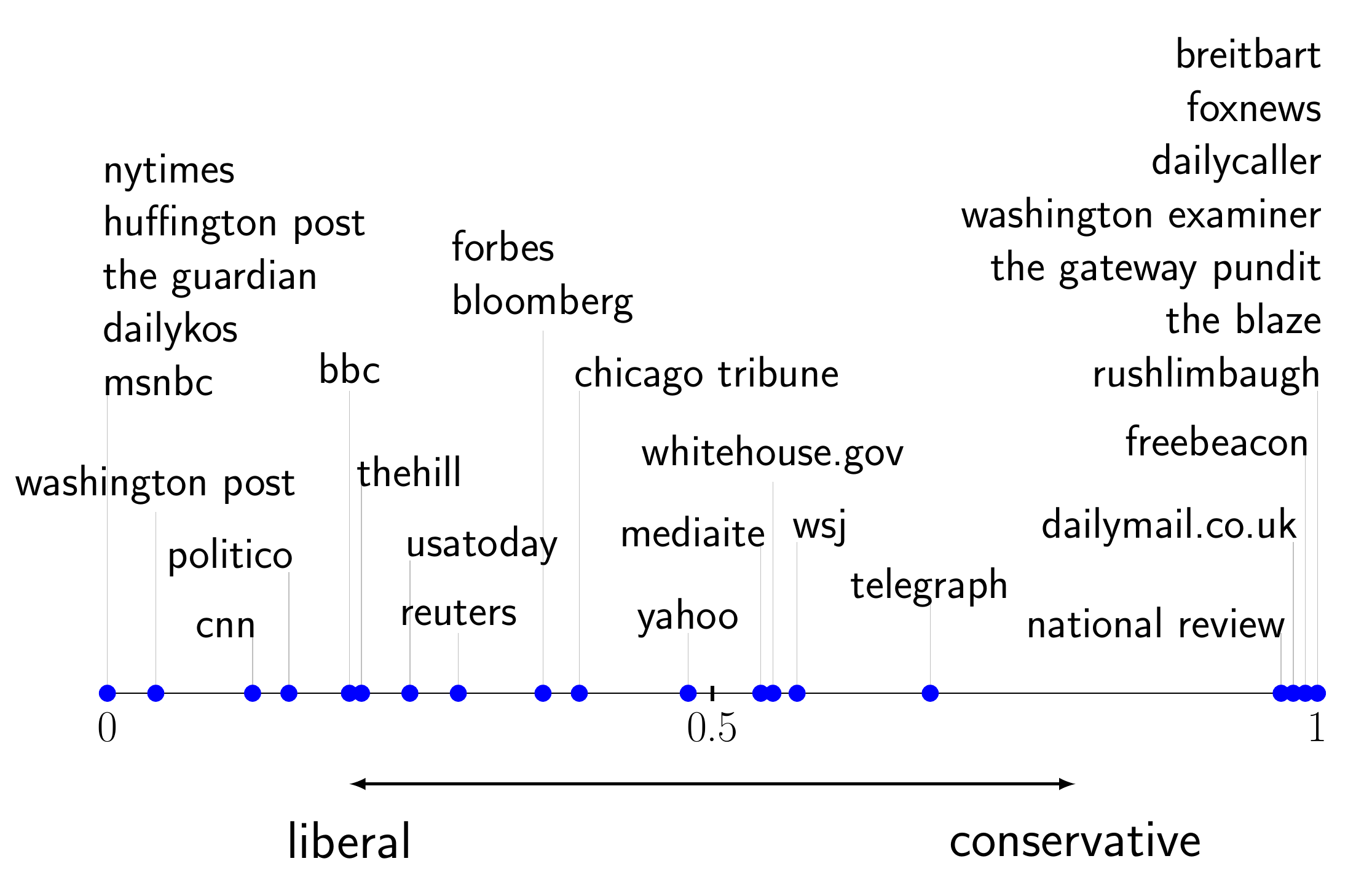}
	\caption{Popular media outlets and their ideology leaning scores computed by our method.}
	\label{fig:popular_user_ideology}
\vspace{-\baselineskip}
\end{figure}

\spara{Ideology estimates for users and sources.}
At a first look, the user ideology scores seem intuitive with the top liberal users being @barackobama (score: 0.0), @berniesanders (0.0), @thedemocrats (0.0) and top conservative users @tedcruz (score: 0.99), @seanhannity (0.99), and @davidlimbaugh (0.9).
Figure~\ref{fig:popular_user_ideology} shows popular news-media outlets
and their ideology leaning scores computed by our method. 
We observe that the position of the news sources is as expected: 
Liberal-leaning news outlets 
(e.g., \emph{nytimes}, \emph{washington post}, \emph{the guardian}) are on the left, and 
conservative news outlets 
(e.g., \emph{fox news}, \emph{breitbart}, \emph{rushlimbaugh}) on the right. 
This is also consistent with the survey-based results found by~\cite{mitchell2014political}. 
While it is easy to identify the extreme left and right, 
it is more difficult to identify the neutral users and sources (like \emph{yahoo}, \emph{mediaite}, \emph{whitehouse.gov}, etc), 
which, in fact, is the most important subset of users and sources 
to tackle the information filter-bubble issue.

\spara{Evaluation of clustering and ideology scores.}
We compare the proposed methods with the baselines on 
(i) quality of ideological cluster separation (\purity, \ARI, \AMI and \NMI) and 
(ii) correlation (\PMCC) between computed ideology score and ground-truth scores. 
The results of both evaluation tasks are listed in Table~\ref{tbl:user-source-eval}. 
Following are some noteworthy observations:

(i) \prob has the best performance among all the methods for both user as well as source clustering. 
We observe that combining network and content information gives consistently best results.
For example, \purity of clustering for combined methods is $20\%$ higher 
than content-only methods for users and $27\%$ higher for source clustering. 

(ii) When comparing ideology scores, \prob performs better than state of the art baselines \kulshrestha and \biaswatch. Note that \biaswatch almost has no correlation with ground truth and performs poorly. This could be due to reasons mentioned in Section~\ref{section:related-work}. Though \kulshrestha works slightly better on the clustering task, it doesn't do so well with the correlation. 
This indicates that our method is able to identify the fine grained ideology scores better, where as \kulshrestha can do better at separating users into binary clusters.


(iii) \NMFSymm, a network-only method, performs quite well for user clustering, 
whereas the results are not satisfactory for source clustering.
Perhaps this can be attributed to the noise in the input matrix $\USmat^T\USmat$, 
caused by, e.g., sparsity and topical diversity of user interests, which is mitigated in \prob because of the joint factorization. 
Similarly, our method performs much better than \retweet and \follow, which do not use both content and network information.
\begin{table}[t]
	\centering
	\caption{Comparison of the proposed method with baseline methods on 
	(i) quality of ideological cluster separation (\purity, \ARI, \AMI and \NMI) and 
	(ii) correlation (\PMCC) between computed ideology score and ground truth. 
	\PMCC for ideology is represented as \PMCCi and for popularity as \PMCCp. 
	Since, we do not have ground truth for popularity of users we do not compute \PMCCp for users. 
	Best results for each measure are marked in bold.}

	\begin{subtable}{1\columnwidth}
		\centering
		\caption{Evaluation for users using \barbera ground truth 
		(upper value in each row) and \avgcontent ground truth (lower value in each row)}
		\label{tbl:user-eval}
	\resizebox{\columnwidth}{!}{
		\begin{tabular}{@{}llllllll@{}}
			\toprule
			Type                                & Method                             & \purity         & \ARI            & \AMI            & \NMI            & \PMCCi       & \PMCCp \\ \midrule
			\multirow{4}{*}{Network Only}      & \multirow{2}{*}{\NMFSymm}          & \textbf{0.928} & \textbf{0.733} & \textbf{0.628} & \textbf{0.629} & \textbf{0.912} & -             \\ 
			&                                    & 0.861          & 0.522          & 0.418          & 0.418          & 0.744          & -             \\ \cline{2-8} 
			& \multirow{2}{*}{\retweet}          & 0.844           & 0.476             & 0.385             & 0.385           & -             & -           \\  &       & 0.839           & 0.461             & 0.395             & 0.399           & -             & -             \\ \cmidrule(l){2-8}
			& \multirow{2}{*}{\follow}             & 0.867          & 0.538          & 0.454          & 0.456          & -             & -             \\ 
			&                                    & 0.845          & 0.476          & 0.382          & 0.382          & -             & -             \\ \hline
			\multirow{4}{*}{Content Only}       & \multirow{2}{*}{\onmft}             & 0.743          & 0.234          & 0.233          & 0.266          & 0.756          & -             \\  
			&                                    & 0.749          & 0.247          & 0.223          & 0.246          & 0.715          & -             \\ \cline{2-8} 
			& \multirow{2}{*}{\dmcc}              & 0.74           & 0.229          & 0.23           & 0.263          & 0.755          & -             \\ 
			&                                    & 0.746          & 0.241          & 0.218          & 0.242          & 0.715          & -             \\ \midrule
			\multirow{8}{*}{Network + Content} & \multirow{2}{*}{\prob}   & \textbf{0.925} & \textbf{0.722} & \textbf{0.62}  & \textbf{0.621} & \textbf{0.904} & -             \\ 
			&                                    & \textbf{0.863} & \textbf{0.528} & \textbf{0.441} & \textbf{0.442} & \textbf{0.772} & -             \\ \cline{2-8} 
			& \multirow{2}{*}{\probngr} & \textbf{0.925} & \textbf{0.722} & \textbf{0.62}  & \textbf{0.621} & \textbf{0.904} & -             \\ 
			&                                    & \textbf{0.863} & \textbf{0.528} & \textbf{0.441} & \textbf{0.442} & \textbf{0.772} & -             \\ \cline{2-8} 
			& \multirow{2}{*}{\kulshrestha}              & \textbf{0.931} & \textbf{0.744} & \textbf{0.637} & \textbf{0.638} & 0.875          & -             \\  
			&                                    & \textbf{0.869} & \textbf{0.547} & \textbf{0.448} & \textbf{0.449} & 0.744          & -             \\ \cmidrule(l){2-8}
			& \multirow{2}{*}{\biaswatch}           & 0.541             & 0.000             & 0.000            & 0.000            & 0.002            & -             \\ 
			& & 0.543            & 0.000           & 0.000             & 0.000             & 0.018             & -             \\ 
	\bottomrule
	\end{tabular}}
	\end{subtable}

\begin{subtable}{1\columnwidth}
	\bigskip
	\caption{Evaluation for sources using \content ground truth}
	\label{tbl:source-eval}
	\centering
\resizebox{\columnwidth}{!}{
\begin{tabular}{@{}llllllll@{}}
	\toprule
	\multirow{2}{*}{Network Only}      & \NMFSymm         & 0.597          & 0.031          & 0.135          & 0.171          & 0.752          & 0.597          \\ \cmidrule(l){2-8} 
	& \follow            & 0.819          & 0.405          & 0.318          & 0.32           & -             & -             \\ \midrule
	\multirow{2}{*}{Content Only}       & \onmft            & 0.606          & 0.039          & 0.145          & 0.181          & 0.746          & 0.593          \\ \cmidrule(l){2-8} 
	& \dmcc             & 0.606          & 0.039          & 0.145          & 0.181          & 0.746          & 0.592          \\ \midrule
	\multirow{4}{*}{Network + Content} & \prob & \textbf{0.826} & \textbf{0.424} & \textbf{0.346} & \textbf{0.348} & \textbf{0.827} & \textbf{0.929} \\ \cmidrule(l){2-8} 
	& \probngr & \textbf{0.822} & \textbf{0.415} & \textbf{0.339} & \textbf{0.341} & \textbf{0.813} & \textbf{0.93}  \\ 
	\bottomrule
\end{tabular}}
\end{subtable}
\label{tbl:user-source-eval}
\vspace{-\baselineskip}
\end{table}

\begin{figure*}[t]
	\begin{subfigure}{0.18\linewidth}
		\centering	
		\includegraphics[scale=0.08]{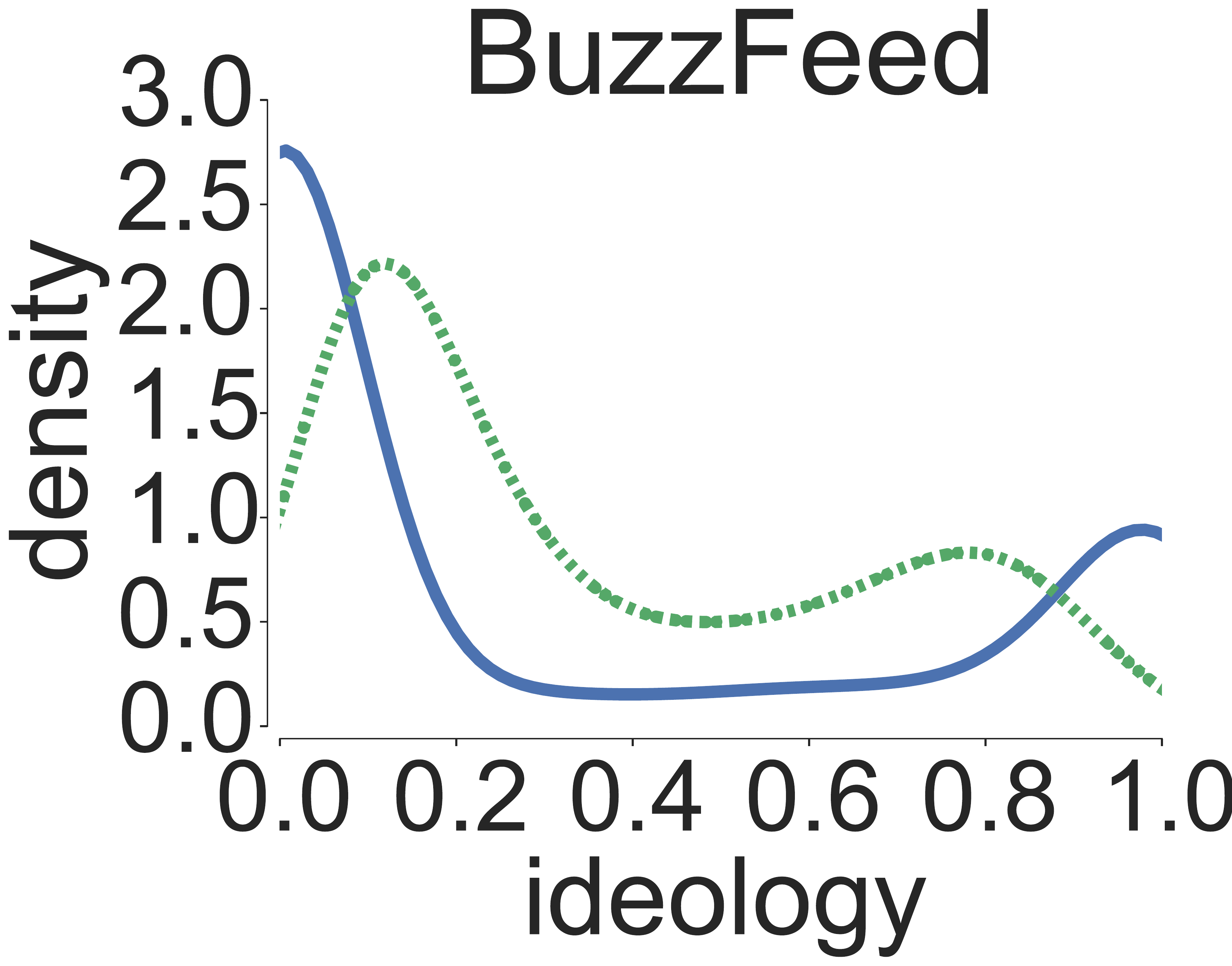}		
	\end{subfigure}
	\begin{subfigure}{0.18\linewidth}
		\centering	
		\includegraphics[scale=0.08]{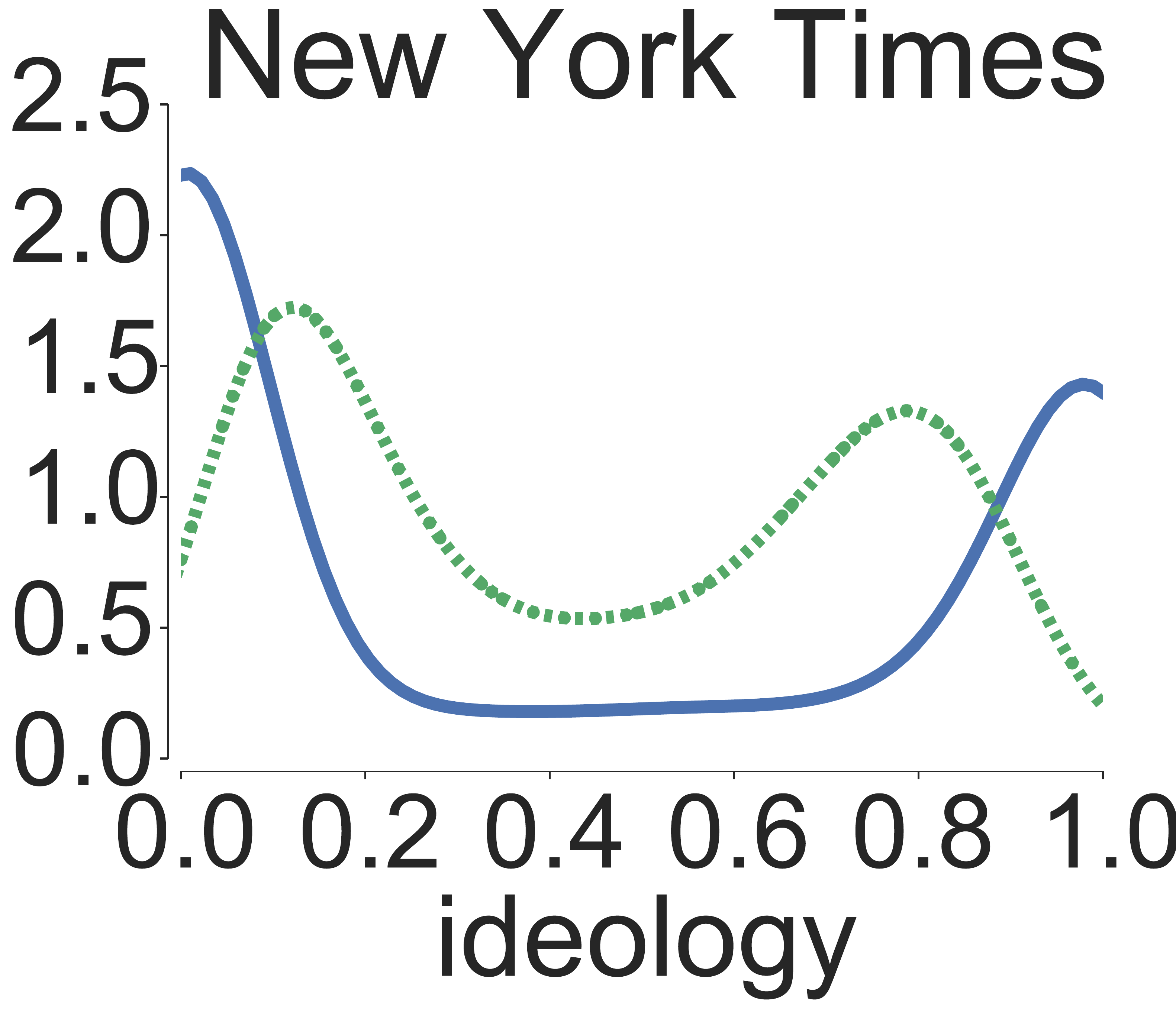}		
	\end{subfigure}
	\begin{subfigure}{0.18\linewidth}
		\centering	
		\includegraphics[scale=0.08]{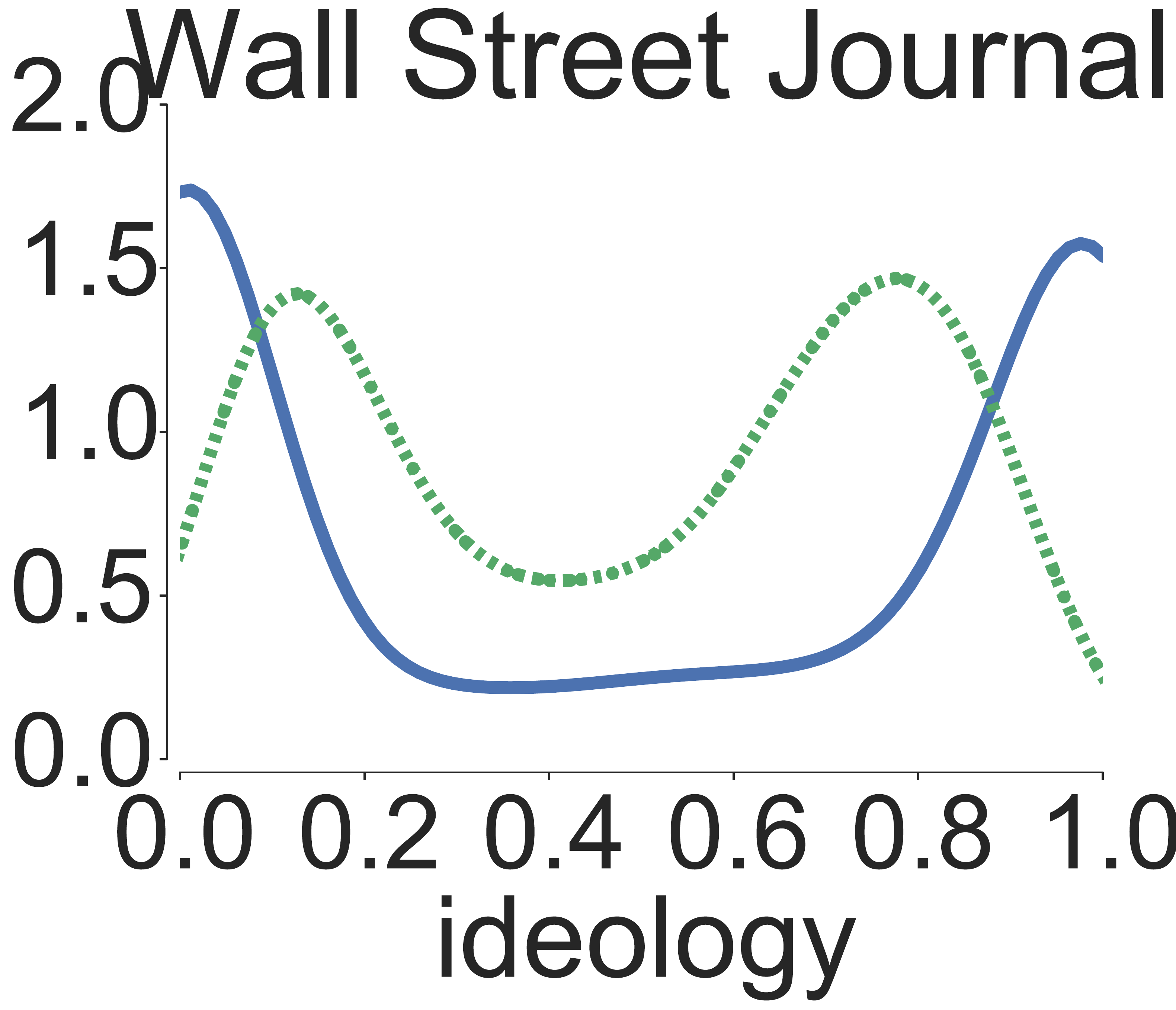}		
	\end{subfigure}
	\begin{subfigure}{0.18\linewidth}
		\centering	
		\includegraphics[scale=0.08]{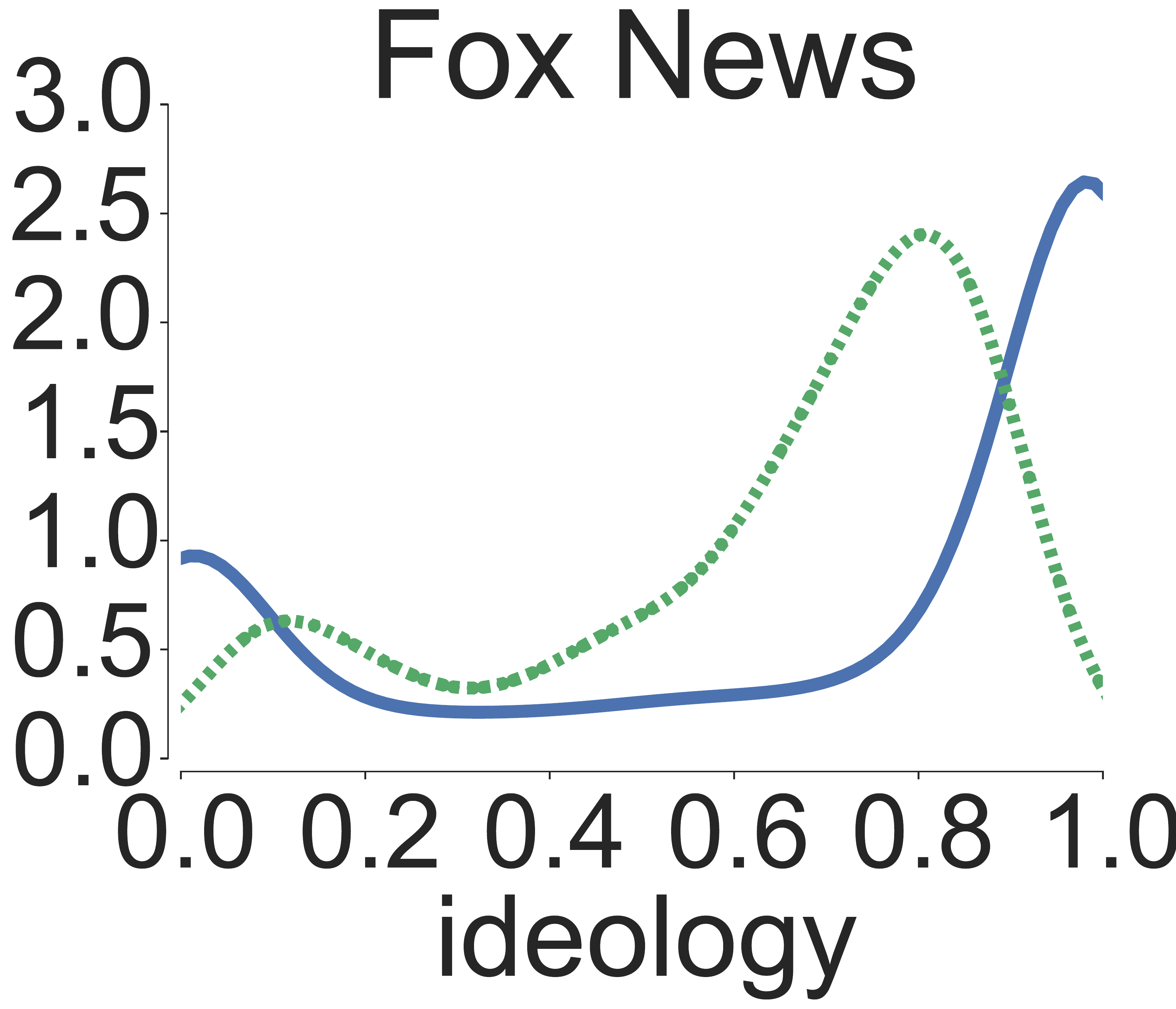}		
	\end{subfigure}
	\begin{subfigure}{0.2\linewidth}
		\centering	
		\includegraphics[scale=0.08]{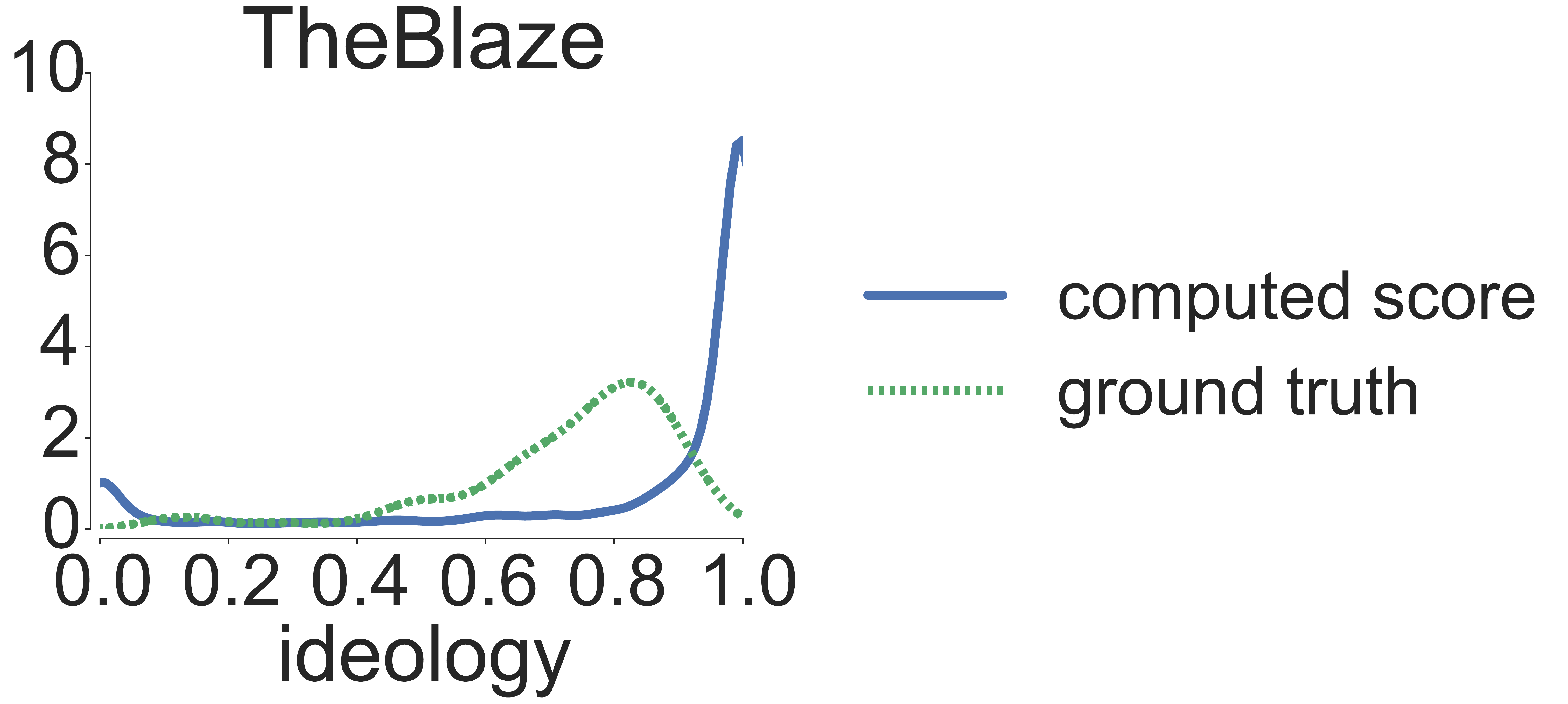}		
	\end{subfigure}
	\caption{Polarization of the audience of news sources.
	Values on the $x$-axis represent the ideology score of users and values on the 
	$y$-axis represent the kernel density estimate of the number of users at each point.}
	\label{fig:user-bin-density}
\vspace{-\baselineskip}
\end{figure*}

\spara{Audience of news sources.}
Figure \ref{fig:user-bin-density} shows 
a kernel density estimate (\kde) of ideology scores computed using the proposed method (solid line) and 
ground truth of ideology scores (dashed line) for all $6\,391$ users 
for a selection of $5$ representative news channels. 
The findings from our experiments are strikingly similar to the results 
computed using extensive user surveys~\cite{mitchell2014political}.
We can clearly observe that there is a non-trivial association between news sources and polarization of users. 

\section{Diffusing the information bubble}
\label{section:case-study}

In the recent years, a few papers~\cite{garimella2017reducing}
have aimed to solve the problem of echo chambers and filter bubbles on social media by connecting users with others outside of their bubble. 
The problem with most of these approaches is that they typically suggest 
technical solutions that do not take into account user biases, such as, 
cognitive dissonance~\cite{festinger1962theory} and biased assimilation~\cite{lord1979biased}.
In addition, in many cases, users themselves are not aware of being present in a bubble, 
due to the non-transparent nature of algorithmic filtering.\footnote{E.g., the filter bubble is claimed to be a reason why many people did not predict correctly the results of Brexit or US elections\\ \url{http://flamingogroup.com/trump-brexit-and-why-we-didnt-see-them-coming}}

To handle both these issues, we propose that a better solution to take users out of their bubble 
is to give them {\em information} and {\em choice}. 
We do this in two steps:
(i) make a user aware of their information bubble, and 
(ii) provide content recommendations that can help users diffuse their bubble.
In this section, we 
demonstrate how to use the latent space representations learned in this paper
to perform both of these steps.

%

\subsection{Visualizing the information bubble} 
\label{sec-ifb-framework}

Recent studies have shown that making users aware of their imbalance in media consumption 
can encourage them to make small yet significant improvements in increasing the diversity 
of their reading~\cite{munson2013encouraging}. 
Inspired by these studies, we suggest providing to the users a visual way to  
explore their information filter bubble.
The way we approach this is by visualizing the user and content in the same space 
to allow users to explore their content consumption. 
To this end, we use the learned latent space to map the users and content in the same space. 
Using their estimated ideological positions (from Section~\ref{section:methodology}), we project users as well as sources in a two dimensional ideology-popularity coordinate space with computed ideology score on $x$-axis and popularity score on $y$-axis. 
Since we do not have popularity score for users, we determine the position of a user on $y$-axis using the average popularity score of the content that the user engages with. 
Since all user scores are on the same scale and relative to each other, we would observe that users and sources with similar ideology lie close to each other. 
Finally, we connect users to the sources that they consume by drawing a link between them. 
The size of a source node is proportionate to the number of times a user has consumed content from the said source. In order to increase the ease of visual interpretation, we color the content according to the ideological learning (blue: liberal, green: neutral and red: conservative). Content not consumed by the user is colored gray. 



\begin{figure}[t]
	\begin{subfigure}{0.505\columnwidth}
		\centering
		\includegraphics[scale=0.185]{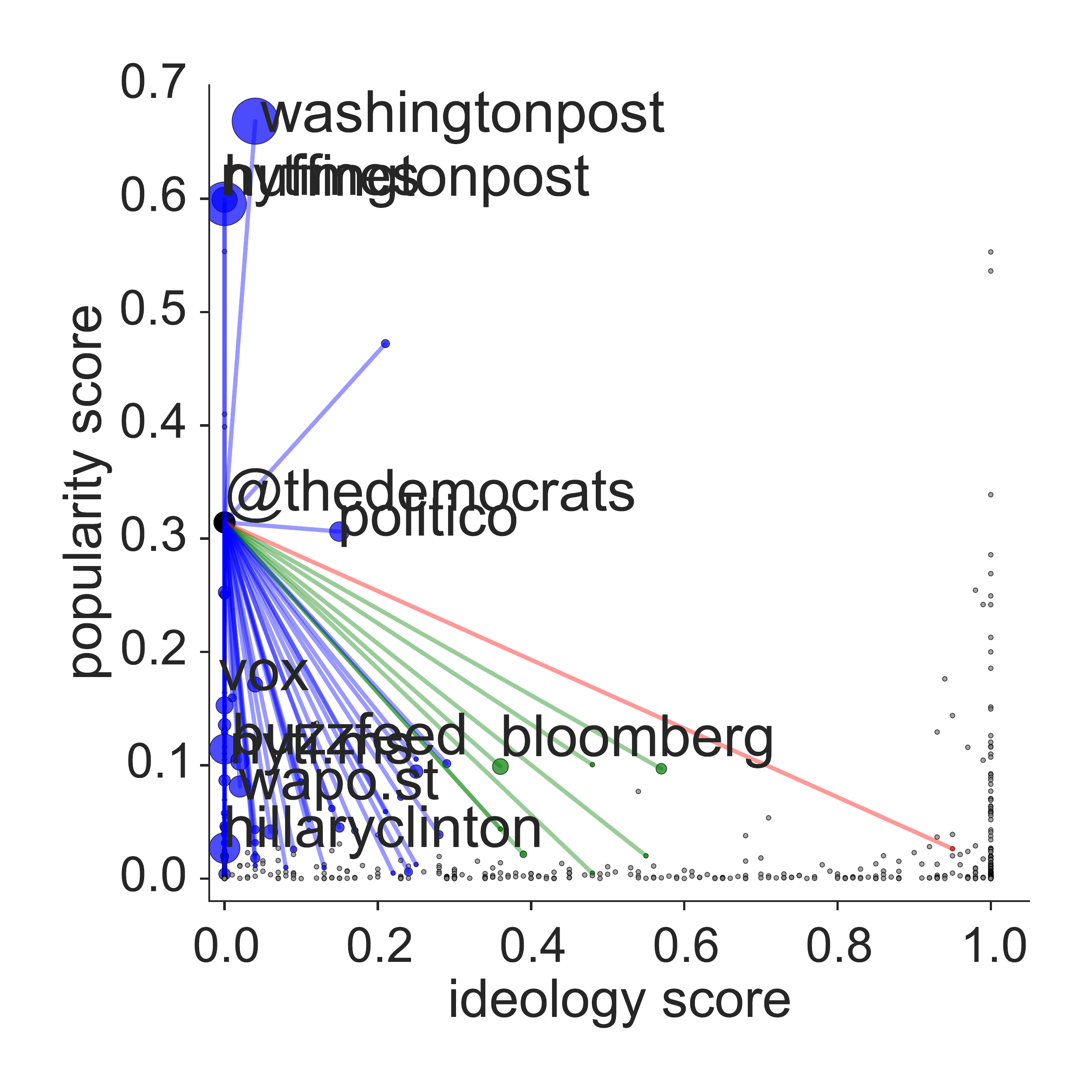}
		\caption{Democratic Party}
	\end{subfigure}
	\hfill
	\begin{subfigure}{0.475\columnwidth}
		\centering
		\includegraphics[scale=0.185]{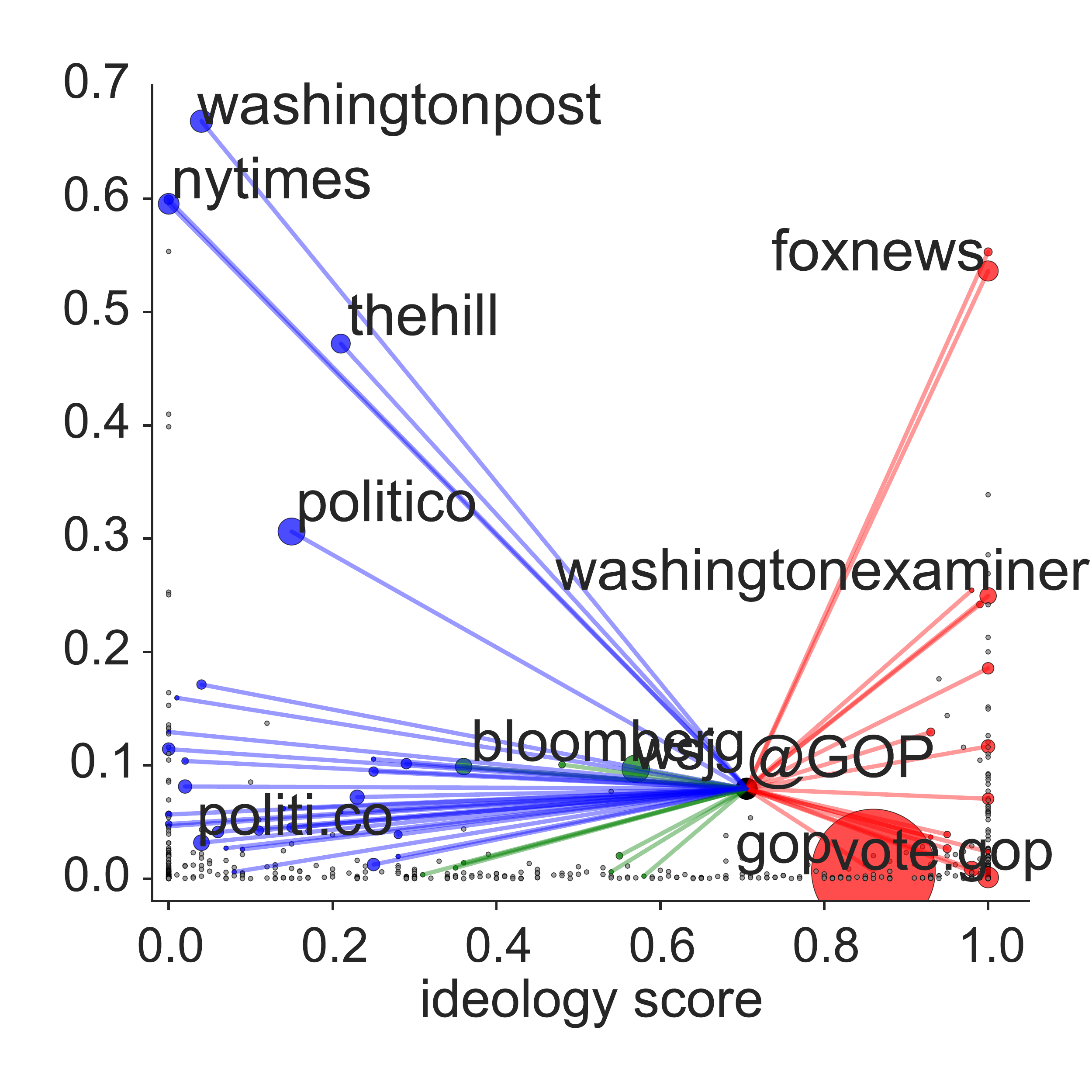}
		\caption{Republican Party}
	\end{subfigure}
	\caption{Ideological position of @thedemocrats and @gop (black dots) and their content engagement. Points in the grey are the sources that the user never interacted with.} 
	\label{fig:prototype-1}
\vspace{-\baselineskip}
\end{figure}

Figure~\ref{fig:prototype-1} presents a prototype for two popular Twitter accounts from the two ends of the political spectrum: 
the Republican Party (@gop) and the Democratic party (@thedemocrats).\footnote{An interactive web version of these plots can be accessed at \\ \url{http://resources.mpi-inf.mpg.de/d5/filterbubble}.}
From this figure, one can visually observe their own ideological positioning as well as the ideology of the content that they engage with. For instance, @thedemocrats is heavily liberal in their ideology (ideology score 0.0). 
The content consumed by @thedemocrats is also heavily biased on the liberal side. As expected, a large fraction of the content they engage with is from the left (mainly liberal media like \emph{nytimes.com} and \emph{washingtonpost.com}), and negligible amount from the opposite point of view, whereas the opposite is true for @gop. 
It is interesting to observe that the Republican party account has a higher engagement with diverse view points than the Democrats. 


\subsection{Making ideologically diverse content recommendations}
\begin{figure}[t]
 	\begin{subfigure}{0.49\columnwidth}
 		\centering
 		\includegraphics[scale=0.34]{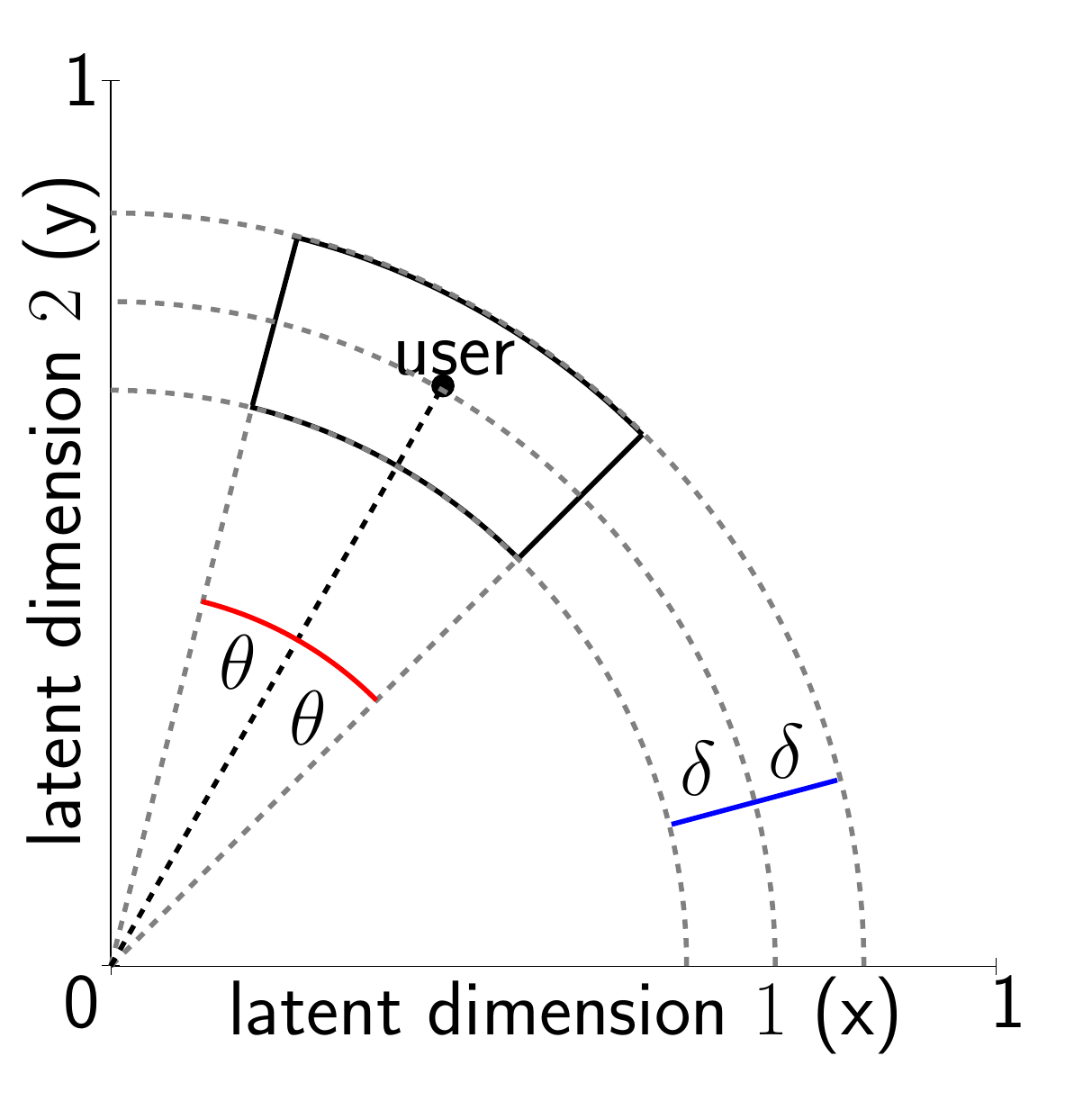}
 		\caption{original latent space}
 		\label{fig:latent-user-polar}
 	\end{subfigure}
 	\begin{subfigure}{0.49\columnwidth}
 		\centering
 		\includegraphics[scale=0.34]{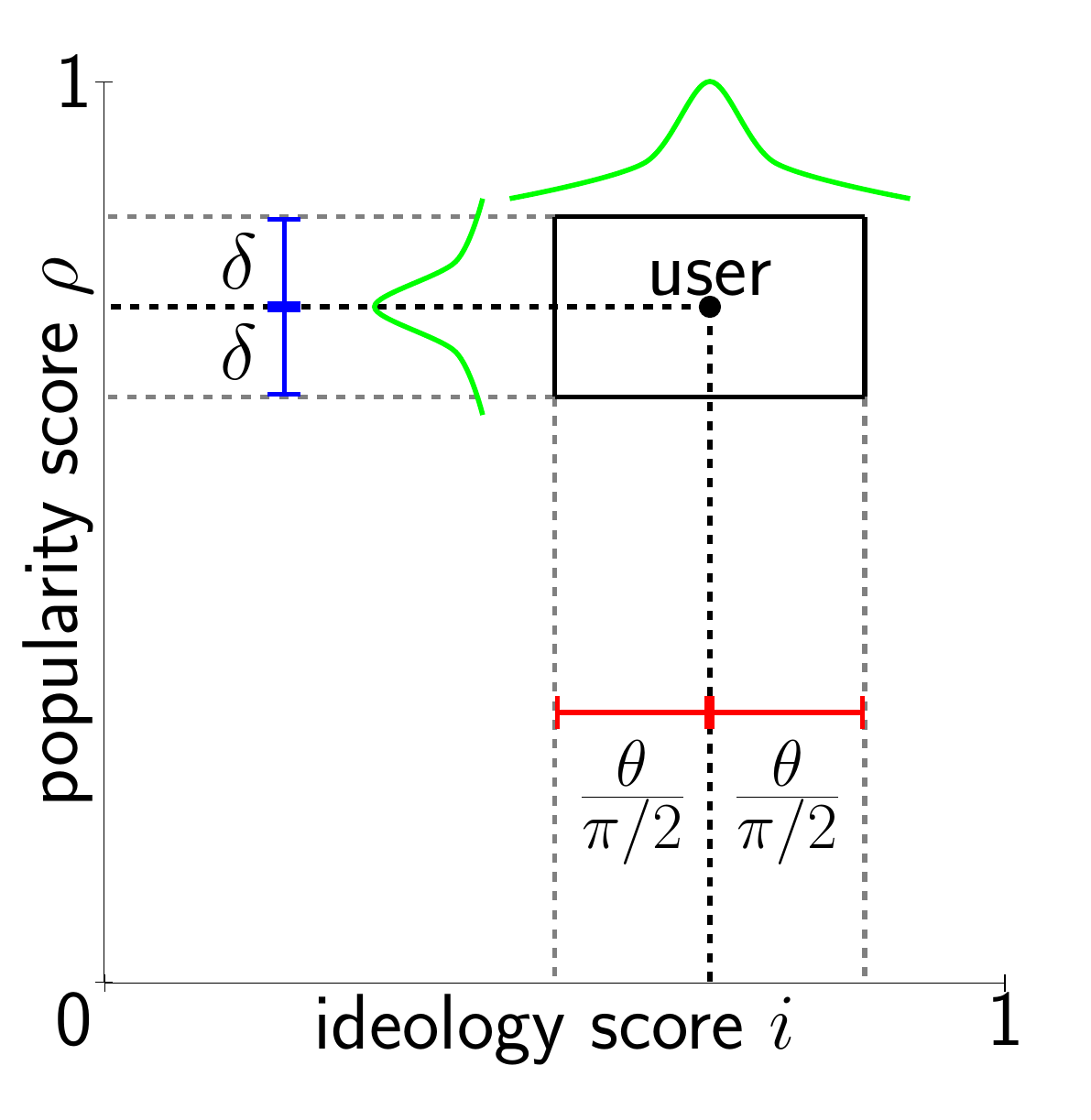}
 		\caption{transformed latent space}
 		\label{fig:latent-user-cartesian-gauss}
 	\end{subfigure}
 	\caption{Logical diagram of user content recommendation by sampling from the Gaussian over ``ideology'' and ``popularity'' positioning.}
 	\label{fig:latents}
\vspace{-\baselineskip}
 \end{figure}

Garimella et al.~\cite{garimella2017reducing} proposed an approach to diffuse a user's filter bubble by connecting him to a user outside his bubble from the opposing viewpoint. Their approach is mainly based on identifying users from opposing sides and optimizing a global function.
Here, we build on top of that idea and use our computed ideology to diffuse a user's bubble by recommending him content from an opposing viewpoint, along with an \emph{option} to choose how willing the user is to explore the other side.
Recommending ideologically diverse content to a user can be controlled 
by the user using two parameters: ideology tolerance threshold $\theta$ and popularity threshold $\delta$. 
Intuitively, a user is more likely to accept content within the region $+\theta$ and $-\theta$ on either side of the user's ideological positioning, and $+\delta$ and $-\delta$ on either side of his popularity position. 
Figure \ref{fig:latent-user-polar} visualizes a hypothetical user in the original ideology latent space and the transformed ideology-popularity coordinate space (detailed in Section~\ref{section:methodology}). 
Consider that we build two Gaussian distributions around the user \emph{box} (see Figure~\ref{fig:latent-user-cartesian-gauss}) with their means centered at user's ideology and popularity score respectively, and variance as a function of the tolerance threshold given as input by the user. 
We can now sample content from these Gaussian distributions and use it for recommending content to the user. 
As desired, in such a sampling, the content close to the user's own ideology and popularity score has a higher probability of being selected. As we move closer to the thresholds, the probability of an article being selected gradually decreases.
This ``{\em box}'' 
gives the space of exploration for a user and depending on the user's willingness to explore (based on parameters $\theta,\delta$), they can see content outside their bubble.

 \cnote[Kiran]{In figure \ref{fig:latent-user-cartesian-gauss} can you change the p to $\varrho$ and in a, change latent l1,l2 to latent dimension 1 (x), latent dimension 2 (y) so that its consistent with figure 1?.}
 \cnote[Preethi]{Done changed p to $\rho$ to be consistent with figure 1.}


\section{Conclusions}
\label{section:conclusions}

We considered the problem of identifying ideological leaning of users and news sources (content) on Twitter. 
The paper tackles two main challenges: (i) learning the ideological latent factors of users and content in a joint model that explores simultaneously user-to-user and user-to-content relations; and (ii) embedding the discovered factors in a common latent space so as to support visualization and exploration of the results. 
%
Our approach distinguishes itself from most existing work in the area in three major ways. 
First, our model aims to learn ideology on a continuous scale rather than a binary \emph{liberal-conservative} opinion, which is a much simpler and well studied problem. Second, our model defines polarization as a multidimensional problem and allows for learning any number of dimensions (ideology, popularity, etc) rather than just one (ideology). 
Third, our model can identify the ideology of both the user and content simultaneously, thus making use of the interdependent structure between content and network.
We also demonstrate how to use the learned latent space in an application of providing tools to visualize a users information bubble and for ideologically diverse content recommendation with the purpose of diffusing their information filter bubble.
%

\noindent
\textbf{Future work.}
One of the main focuses of this work was to create a strong technical foundation to understand the problem of online polarization.
We believe that the 
methods presented in this paper provide several avenues for future work 
in multiple emerging interdisciplinary research areas, 
for instance, humanly interpretable and explanatory machine learning, transparent recommendations and a new era of social media platforms that encourage discussion and debates between users of diverse view points, thus helping to reduce the ideological segregation of users instead of reinforcing~it.


\noindent
\textbf{Acknowledgments.}
This work was supported by 
the Academy of Finland projects ``Nestor'' (286211) and ``Agra'' (313927), 
and the EC H2020 RIA project ``SoBigData'' (654024).

\bibliographystyle{abbrv}
\bibliography{references}

\end{document}